\documentclass[11pt, a4paper]{article}
\pdfoutput=1

\usepackage{jinstpub}

\usepackage{lineno}
\usepackage{soul}
\usepackage{ulem}
\usepackage{multirow}
\graphicspath{{./figures/}}
\usepackage{color}

\usepackage{chngcntr}
\counterwithout{equation}{section} 

\title{Qualification Tests of the R11410-21 Photomultiplier Tubes for the XENON1T Detector}

\author[a]{P.~Barrow,}
\author[a]{L.~Baudis,}
\author[b]{D.~Cichon,}
\author[b]{M.~Danisch,}
\author[a]{D.~Franco,}
\author[b]{F.~Kaether,}
\author[a]{A.~Kish,}
\author[b]{M.~Lindner,}
\author[b]{T.~Marrod\'an~Undagoitia,}
\author[a,*]{D.~Mayani,%
\note[*]{Corresponding authors.}}
\author[b,*]{L.~Rauch,}
\author[a]{Y.~Wei,}
\author[a]{J.~Wulf}
\affiliation[a]{Physik Institut, Universit\"{a}t Z\"{u}rich,\\
8057 Z\"{u}rich, Switzerland}
\affiliation[b]{Max-Planck-Institut f\"{u}r Kernphysik\\
69117 Heidelberg, Germany}

\emailAdd{dmayani@physik.uzh.ch}
\emailAdd{rauch@mpi-hd.mpg.de}

\abstract{The Hamamatsu R11410-21 photomultiplier tube is the photodetector of choice for the XENON1T dual-phase time projection chamber. The device has been optimized for a very low intrinsic radioactivity, a high quantum efficiency and a high sensitivity to single photon detection. A total of 248 tubes are currently operated in XENON1T, selected out of 321 tested units. In this article the procedures implemented to evaluate the large number of tubes prior to their installation in XENON1T are described. The parameter distributions for all tested tubes are shown, with an emphasis on those selected for XENON1T, of which the impact on the detector performance is discussed. All photomultipliers have been tested in a nitrogen atmosphere at cryogenic temperatures, with a subset of the tubes being tested in gaseous and liquid xenon, simulating their operating conditions in the dark matter detector. The performance and evaluation of the tubes in the different environments is reported and the criteria for rejection of PMTs are outlined and quantified.}


\begin{document}

\maketitle
\flushbottom

\section{Introduction}
The direct detection of dark matter particles scattering off target nuclei is one of the most sought-after measurements in modern physics. In the last decade, dual-phase time projection chambers (TPC), operated with liquefied xenon (LXe), have reached the highest sensitivity for Weakly Interacting Massive Particle (WIMP) interactions above masses of $5\,\rm{GeV}/c^2$~\cite{Undagoitia:2015gya}\cite{Baudis:2016qwx}. The XENON1T experiment~\cite{Aprile:2015uzo} is designed to further improve this sensitivity, searching for WIMPs through their scattering off xenon nuclei. At the expected recoil energies of a few keV, the recoiling nucleus can create excimer molecules in xenon which, upon dissociation, produce scintillation of vacuum ultraviolet (VUV) photons with a wavelength of 178\,nm~\cite{Cheshnovsky1972}. In addition, particle recoils produce free electrons that are drifted and extracted from the liquid by an applied electric field to be subsequently amplified in the gaseous xenon volume. 
The amplified electron showers 
produce a secondary scintillation signal through the process of electroluminescence\,\cite{Lansiart1976}. It becomes evident that a successful detection of particle interactions in liquid xenon can only be achieved by an efficient detection of VUV photons. To reach the highest possible sensitivities to dark matter interactions, the photomultiplier tubes (PMTs) are required to have a high photon detection efficiency (quantum efficiency) at a wavelength of 178\,nm, ultra low radioactivity levels~\cite{Aprile:2015lha} and stable long-term performance at operating temperatures of $-100$\,$^{\circ}$C. 

In this article, the evaluation tests of 321 Hamamatsu R11410-21 PMTs are described. The selection criteria for the final 248 tubes for XENON1T are discussed, based on characteristic parameters such as the dark count rate, quantum efficiency, afterpulse rate, light emission, long-term stability, peak-to-valley ratio and single photoelectron (SPE) resolution. The general testing procedure and experimental facilities are introduced in section~\ref{sec:exp}. The distributions of measured and derived PMT parameters are shown in section~\ref{sec:res}. Section~\ref{sec:lightemission} describes the emission of light by some of the tested PMTs, being one of the major rejection criteria. Dedicated measurements of the long-term stability in cryogenic xenon environments are presented in section~\ref{sec:xetests}. Here the evolution of the dark count rates and gains is reported. Section~\ref{sec:ap} describes a detailed study of the afterpulse spectra and the methods used to determine the presence of leaks in faulty PMTs by identification of residual gas molecules. The conclusions of this article are summarized in section~\ref{sec:sum}. 

Complementary studies of earlier versions of the R11410 phototube have been reported in~\cite{Lung:2012pi}\cite{Baudis:2013xva}. This PMT is also operated in experiments using liquid and gaseous xenon, such as PandaX\,\cite{Cao:2014jsa}, NEXT\,\cite{Alvarez:2012flf} and RED\,\cite{Akimov:2012aya}\cite{Akimov:2015aoa}, while a version for operation in liquid argon is used in DarkSide\,\cite{Agnes:2014nla} and GERDA\,\cite{Agostini:2015boa}. The long term performance of 37 R11410-MOD PMTs within the PandaX dark matter experiment is presented in\,\cite{Li:2015qhq}.
 
\section{Testing procedure and experimental setups} \label{sec:exp}

The Hamamatsu R11410-21 is a circular quartz-windowed PMT with a 3-inch diameter produced specifically for low background applications, such as dark matter detectors. The final version of the tube is a result of several iterations of construction material selection for the reduction of the total radioactivity of the device. After production and shipment, each PMT has been screened for its intrinsic radioactivity with high-purity germanium detectors~\cite{Baudis:2011am}\cite{Budjas:2007yj}\cite{Heusser:2004}\cite{Neder:2000}. The average levels measured per PMT are lower than 13\,mBq/PMT for $^{238}$U and 0.4\,(0.1)\,mBq/PMT for $^{228}$Th~\cite{Aprile:2015lha}.

After screening, every PMT is tested in air at room temperature and in nitrogen gas at around $-100\,^{\circ}$C, the same temperature as when operated in liquid xenon (which will also be referred to as ``cold''). The testing setups used for all PMTs are described in section~\ref{sec:mpik_setups}. These overall tests verify the agreement of various PMT parameters with the production specifications, such as the gain, transit time spread, dark count and afterpulse rates. In addition, a PMT  characterization is performed, including the voltage dependency of the gain, the signal resolution and the afterpulse spectrum. Every PMT is cooled down a minimum of 2 times to evaluate its performance at cryogenic temperatures and test its level of light emission (some PMTs have shown a quantifiable emission of photons, which will be described in section~\ref{sec:lightemission}). 

A subset of the tubes has been selected to be further tested in gaseous and liquid xenon with the setup described in section~\ref{sec:mxl_setup}. There, the PMTs are operated over longer periods of time, from a couple of weeks to several months, and the dark count rate and gain evolution are measured. Dedicated measurements are performed to study the afterpulse spectra after each cool down.

The voltage divider bases for the PMTs in these setups have the same design as those used in XENON1T, with a negative bias operation. The divider has been developed for an optimal PMT performance in terms of gain and linearity, as well as the restrictions in heat dissipation and radioactivity for a LXe dark matter detector~\cite{Annika:thesis}.    

\subsection{General testing facility} \label{sec:mpik_setups}

The first experimental setup, at the Max-Planck-Institut f\"ur Kernphysik (MPIK) in Heidelberg, is a dark room designed to test 12 PMTs simultaneously at room temperature. Installed within a large Faraday cage, which provides a low-noise environment, each PMT is mounted in a compartment equipped with an optical fiber connected to an LED emitting at 380\,nm. The LED is controlled via a custom VME board providing a narrow pulse width of 1.4\,ns for precision timing measurements. This setup allows to perform a high voltage scan to determine the gain values, measure the single photoelectron response, the afterpulse spectrum and transit time spread of the PMTs~\cite{Bauer:2011ne}.

\begin{figure}[h]
  \begin{center}
   \includegraphics[angle=0,width=0.9\textwidth]{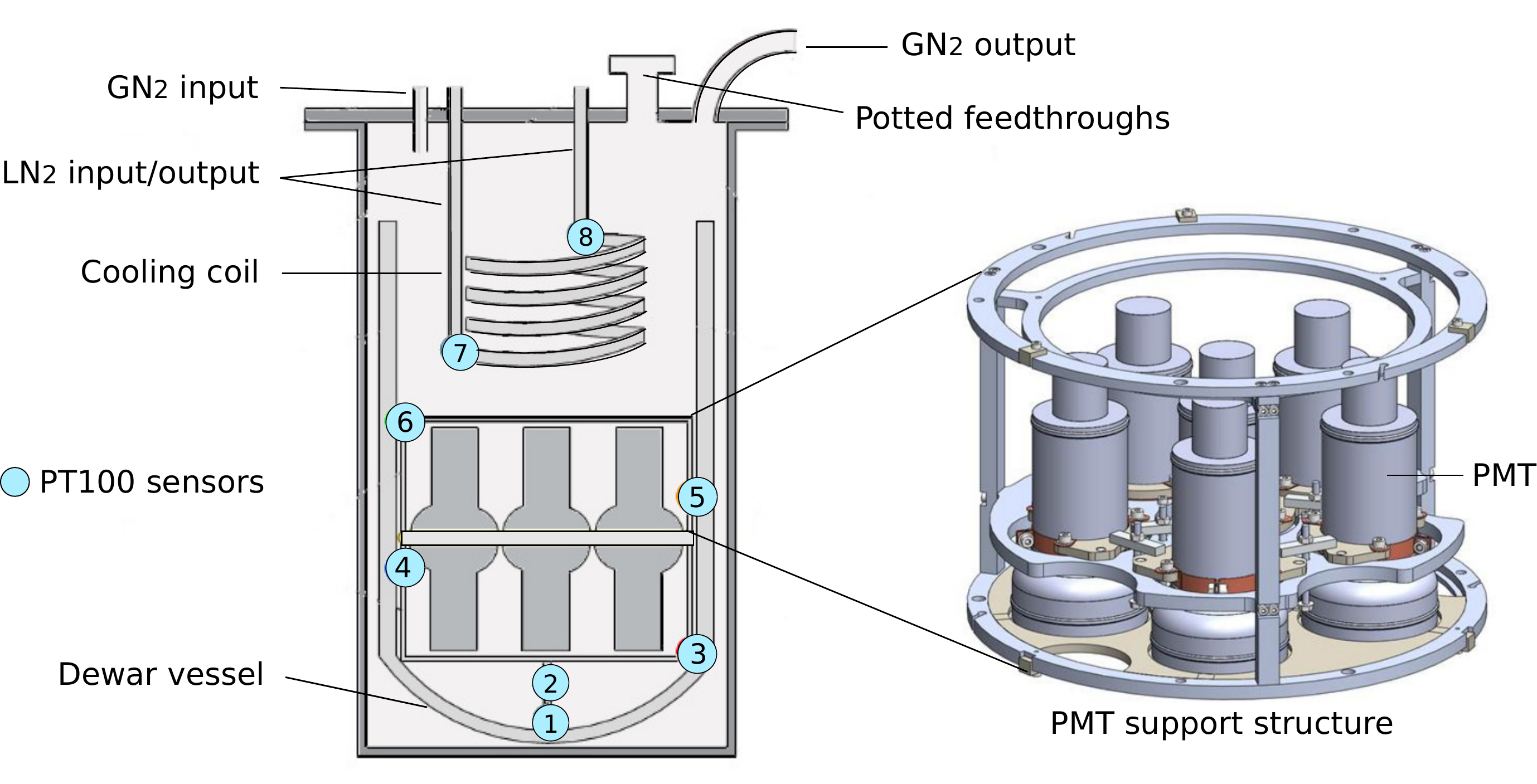}
   \caption[]{Schematic of the PMT setup for cryogenic temperature testing in nitrogen gas stabilized  at $-100\,^{\circ}$C. Two arrays of 6 PMTs each (left figure) are inserted into the dewar with the PMT windows facing the opposite array. Several PT100 sensors allow to measure and control the inner temperature within a few degrees by regulating the flow of the nitrogen through the coil.  \label{Fig:setup}}
  \end{center}	
\end{figure}

The second setup, also at MPIK, consists of a cooling tank designed for stable operation at cryogenic temperatures. The inner volume is permanently flushed with nitrogen vapor, with the system open to the ambient atmosphere. Cooling is provided by a copper coil at the top of the vessel, constantly flushed with liquid nitrogen. Inside the tank, two parallel arrays of 6 PMTs each are operated with the tube windows of one array facing those of the other at a distance of 2\,cm. The temperature is measured at several positions with PT100 sensors to ensure a stable temperature within a few degrees during the measurements. A schematic of this setup is shown in figure~\ref{Fig:setup}.

Each PMT signal is initially amplified by a factor of 10 and sent to a fan out. The signal is split and read out by a charge-to-digital converter (QDC). The signal is amplified once more by a factor 10 and sent to a discriminator to be subsequently analyzed by a scaler and a time-to-digital converter (TDC). The TDC has a maximal time span of $1.2\,\mu s$ with a time resolution of $0.3$\,ns. The outputs of the QDC, TDC and scaler are read out by a computer where the time information of the LED trigger is also available to, for example, control the QDC time-integration window. Detailed information can be found in~\cite{Bauer:2011ne}.


\subsection{Xenon testing facility} \label{sec:mxl_setup}

A subset of the PMTs (around 15\,\%) has been tested in a third experimental setup, located at the Physik-Institut of the University of Zurich, where they are operated in both gaseous and liquid xenon for periods ranging from  a couple of weeks up to several months. These tests evaluate the performance and stability of the PMTs in the same conditions in which they are operated in the XENON1T dual-phase TPC, with the top array in gas and the bottom array in liquid.

\begin{figure}[h]
  \begin{center}
   \includegraphics[angle=0,width=0.65\textwidth]{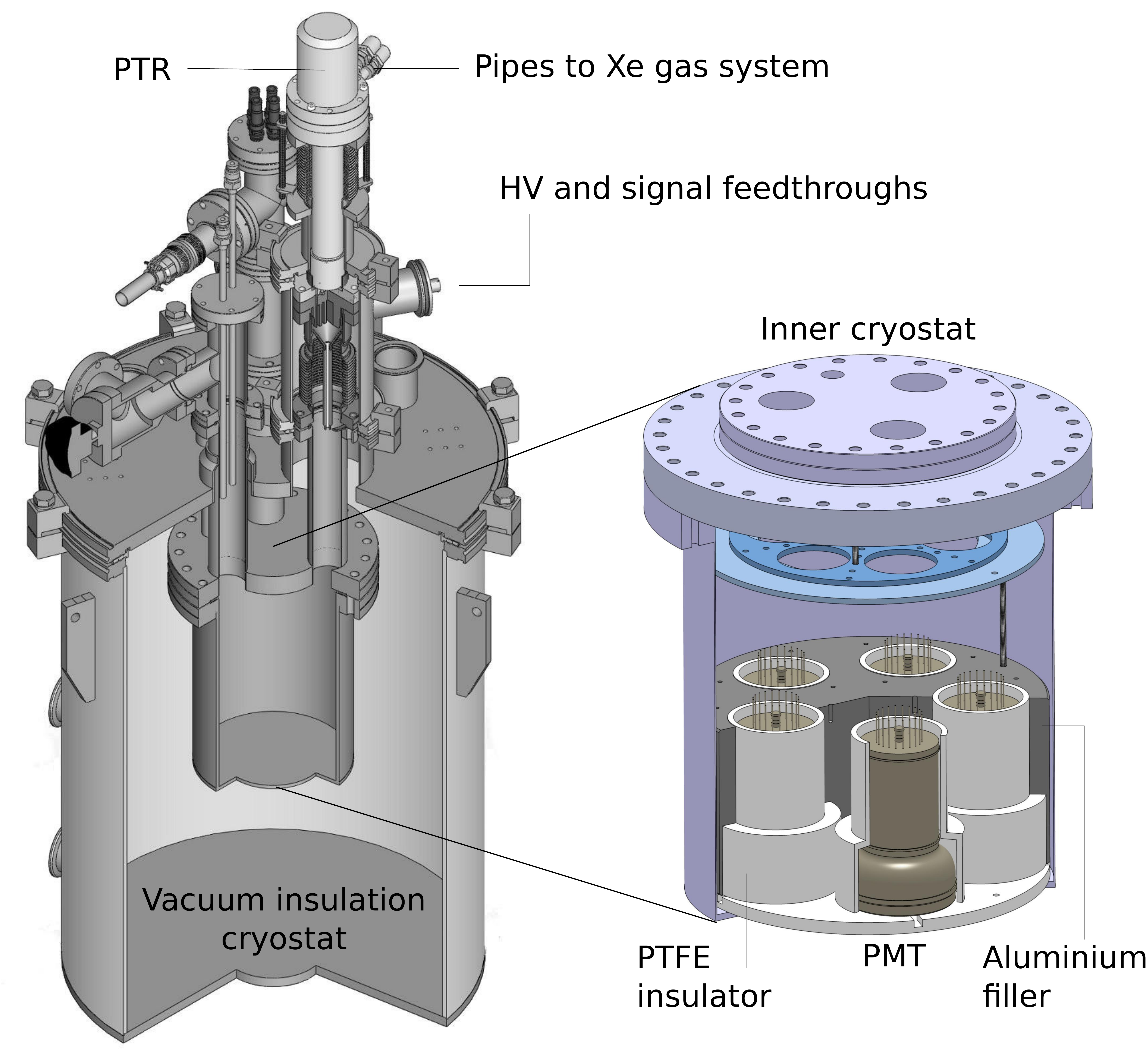}
   \caption[]{Schematic of the testing facility for PMTs in both gaseous and liquid xenon environments. The inner cryostat houses up to 5 PMTs simultaneously. The aluminium filler reduces the amount of LXe required to cover the PMTs, while the PTFE covers insulate the metallic bodies. A larger cryostat provides the inner chamber with vacuum insulation during cryogenic operation. \label{Fig:marmotxl}}
  \end{center}	
\end{figure}

The xenon testing facility consists of a double-walled vacuum-insulated cryostat housing up to 5 R11410 PMTs simultaneously. Figure~\ref{Fig:marmotxl} shows a schematic of the experimental setup. The inner cryostat contains 5 PMTs facing a polytetrafluoroethylene (PTFE) reflector. A circular space is provided below each PMT and linked to all others through a central volume of the reflector. A PTFE insulator separates the PMTs from the aluminium filler, used to reduce the amount of LXe required to cover the sensors. Two temperature sensors have been installed, one at the bottom next to the PMT windows and another above the filler at the height of the PMT pins. A blue LED (470\,nm wavelength) is installed in the common volume of the PTFE reflector between the PMTs.

\noindent In addition, four optical fibers are placed at the same location, coupled to an external LED.   

Gaseous and liquid xenon can be filled into the system, surrounding the PMTs through the space between the bodies and the PTFE covers. The cooling power for liquefaction is provided by a pulse tube refrigerator (PTR). Xenon is circulated through a dedicated gas system that controls the flow into and out of the cryogenic chamber, with a heated getter that removes impurities. The setup features the same 1.83\,mm PTFE coaxial cables and single wire Kapton cables currently used for signal and HV in the PMT arrays of XENON1T, respectively. Custom connectors designed with low radioactivity materials specifically for XENON1T have also been used and tested here~\cite{Gaudenz:thesis}. Potted feedthroughs link the cabling of the PMTs inside the vessel to the external electronics for data acquisition. The signals go through a ten-fold amplifier and a fan out. For counting measurements, such as dark count rates, a discriminator is used along with a multi-channel scaler. Signals are digitized with a flash ADC and analyzed with a dedicated software.


\section{PMT parameter measurements and distributions} \label{sec:res}

This section summarizes the tests performed on 321 PMTs, of which 248 were selected for XENON1T. The measurements have been performed in the testing facility described in~\ref{sec:mpik_setups}, with every PMT following the same procedure for comparable results. The derived parameters for the PMT characterization are briefly defined and their implications for a dark matter search are outlined. 

\subsection{Quantum efficiency}

To enable a low energy threshold of the XENON1T experiment a high light collection efficiency is required and, hence, a high quantum efficiency (QE) of the PMTs at a wavelength of 178\,nm. The tubes have been pre-selected, as contracted, with a QE higher than 28\% measured by the producer at 175\,nm, which is within the xenon scintillation peak.

\begin{figure}[h!!]
  \begin{center}
   \includegraphics[angle=0,width=0.52\textwidth]{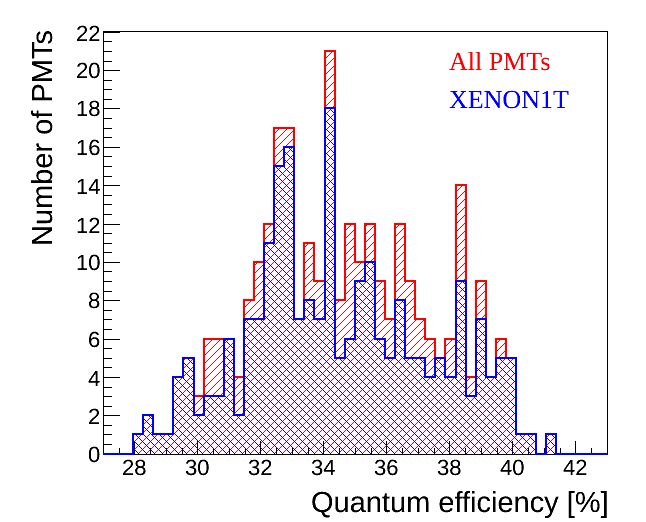}
 \end{center}	
  \caption[]{Quantum efficiency at a wavelength of 175\,nm for all tested PMTs (red) and the 248 selected for XENON1T (blue), as measured by Hamamatsu. The company pre-selected the PMTs with a QE higher than 28\%, as contracted. Both distributions show an average around 34.5\,\% and a standard deviation of 2.8\,\%. \label{Fig:qe}} 
\end{figure}

The QE distributions for all PMTs and the XENON1T selection, as measured by Hamamatsu, are shown in figure~\ref{Fig:qe}. The average QE is considerably high, with a mean value of approximately 34.5\,\% and a standard deviation of 2.8\,\% for both distributions. A study of the spatial distribution of the photon detection sensitivity can be found in~\cite{Baudis:2015rza}, while the temperature and wavelength dependence of the QE has been studied in~\cite{Lyashenko:2014rda}. Furthermore, the possibility of a double photoelectron emission in the cathode due to UV light is non-negligible and must be accounted for in a dark matter experiment using liquid xenon~\cite{Faham:2015kqa}.

\subsection{Gain distribution}
\begin{figure}[h!]
  \begin{center}
  \makebox[\textwidth][c]{
   \includegraphics[angle=0,width=0.50\textwidth]{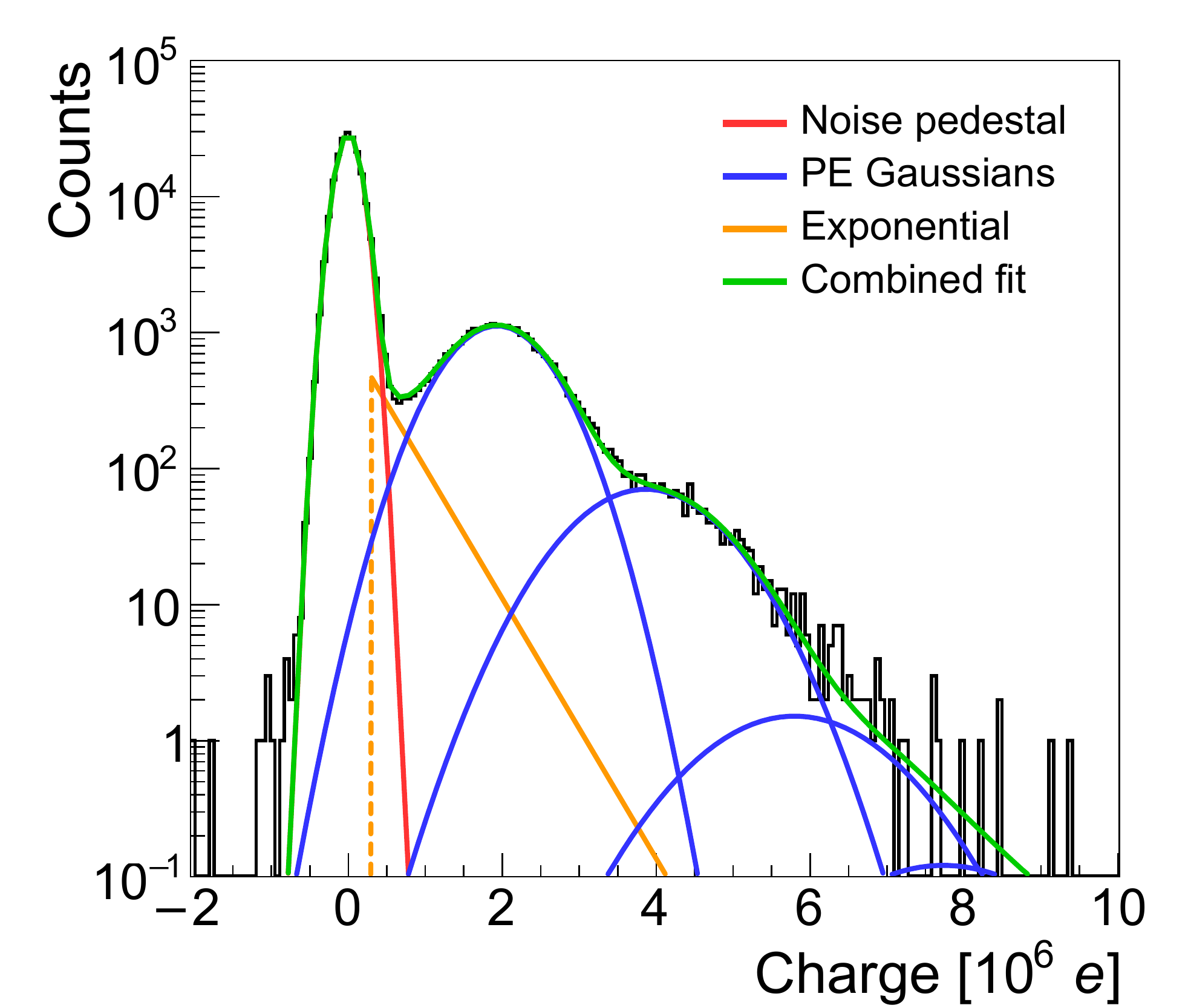}
   \includegraphics[angle=0,width=0.52\textwidth]{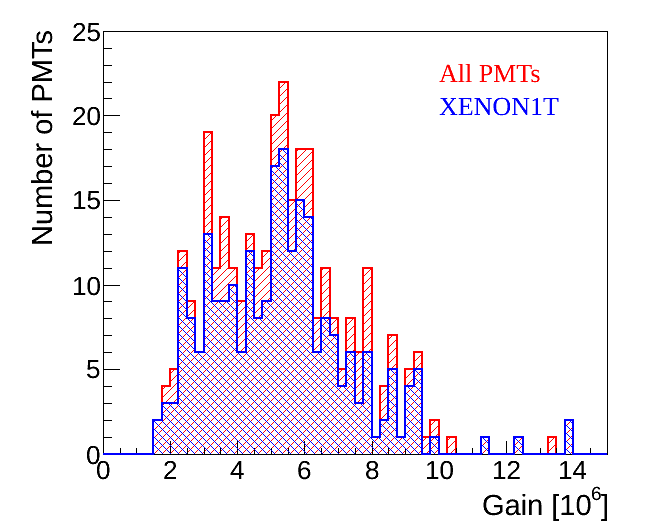}
   }
   \caption[]{(Left) PMT single photoelectron spectrum and its various components. The pedestal (red) corresponds to the noise distribution. The Gaussians for the SPE and multi-PE peaks are shown in blue. The exponential function (yellow) is included to account for the under-amplified photoelectron signals and improves the overall fit (green). (Right) Distribution of the measured gains at 1500\,V for the 248 selected tubes (blue) and all the 321 tested PMTs (red). An average value of $5.3\times 10^6$ with a standard deviation of $2.1\times 10^6$ is computed for the selected group and a similar value is found for all PMTs: $5.4\times 10^6$ with a standard deviation of $2.1\times 10^6$.  \label{Fig:SPE}}
  \end{center}	
\end{figure}

The PMT response to a light source inducing single photoelectrons can be quantified by measuring the charge spectrum obtained using the QDC (as described in section~\ref{sec:mpik_setups}). The result is shown in figure~\ref{Fig:SPE}~(left). 

The baseline noise from the electronics results in a pedestal in the SPE spectrum described by a Gaussian of amplitude $A_0$, mean $\mu_0$ and width $\sigma_0$. The pedestal is followed by the SPE peak with Gaussian parameters $A_1$, $\mu_1$ and $\sigma_1$. Given the single photon intensity of the light source, higher PE contributions follow Poissonian statistics and their Gaussian parameters are thus dependent on the SPE values as shown in the total fit function written as a function of the measured charge $x$:
 
\begin{equation} \label{eq:1}
f(x) = A_0\exp\left(-\frac{(x-\mu_0)^2}{2\,\sigma_0^2}\right) + \sum_{i=1}^N A_i \exp\left(-\frac{(x-i\mu_1)^2}{2i\sigma_1^2}\right) + B\exp(-x\tau).
\end{equation}

\noindent The exponential function is included to account for the under-amplified photoelectron signals and its addition improves the overall fit at the valley between the pedestal and single photoelectron peak (as in~\cite{Bauer:2011ne} and~\cite{Bellamy:1994bv}). The gain is computed as $g = \frac{\mu_1}{e} $, with $e$ being the elemental charge.

Figure~\ref{Fig:SPE}~(right) shows the distribution of gains derived from the SPE spectra at 1500\,V (for simplicity, the absolute HV value is stated through this article, although the PMT bias voltage is always negative). The average gain for all PMTs at 1500\,V is $ 5.4 \times 10^6$ with a standard deviation of  $2.1\times 10^6$, while the value for those selected for XENON1T is $5.3 \times 10^6$ and a standard deviation of  $2.1\times 10^6$. 

The gain at room temperature has been measured for every PMT at 8 different voltages between 1320\,V and 1680\,V with an LED illumination at single-photon intensity. The scan allows to determine the gain as a function of HV for each PMT, fitting the data with a power law curve. 
The optimal gain selection in XENON1T will consider such parameters as the SPE resolution and peak-to-valley ratio, described in the following section.

\subsection{Single photoelectron resolution and peak-to-valley ratio}
The SPE peak resolution is defined as $R = \frac{ \sigma _1}{\mu_1}$, where $\sigma _1$ and $\mu _1$ are the standard deviation and the mean of the Gaussian function describing the SPE peak (see equation \ref{eq:1}). In turn, the peak-to-valley ratio is calculated by dividing the maximum height of the SPE peak and the minimum value of the valley between the peak and the noise pedestal. 

\begin{figure}[h!]
  \begin{center}
  \makebox[\textwidth][c]{
   \includegraphics[angle=0,width=0.52\textwidth]{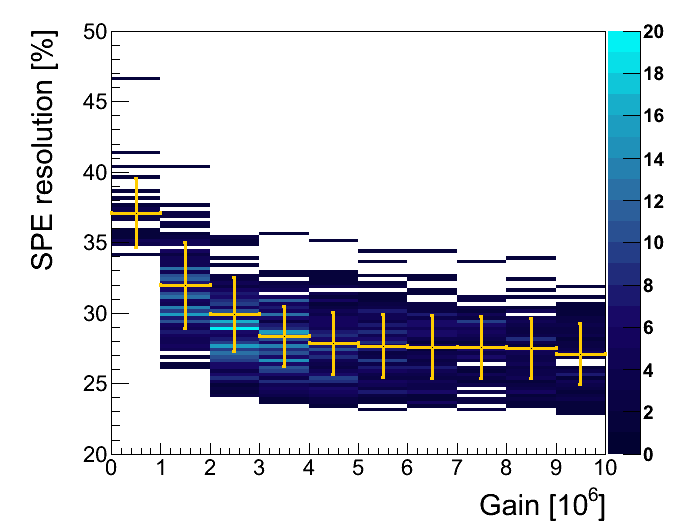}   
   \includegraphics[angle=0,width=0.52\textwidth]{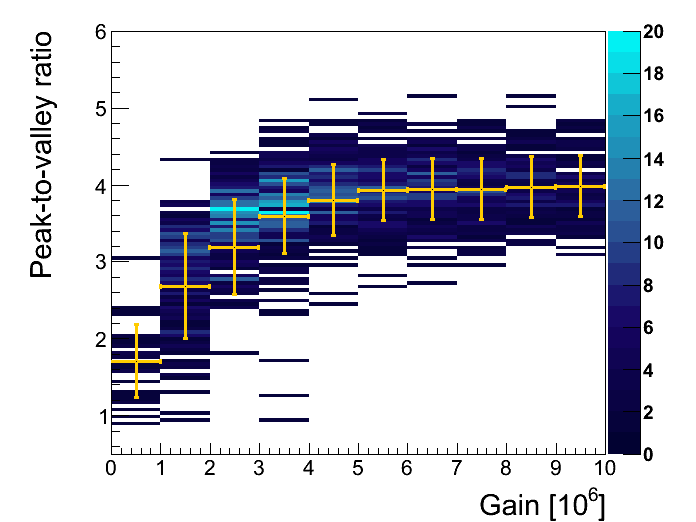}
   }
   \caption[]{(Left) Single photoelectron resolution, measured at room temperature, as a function of gain for the XENON1T selected PMTs. (Right) Peak-to-valley ratio as a function of gain. The color coded scale displays the number of PMTs. For each bin the average resolution and standard deviation are shown by the yellow markers.\label{Fig:spe_quant}}
  \end{center}	
\end{figure}

Both parameters can be extracted from the measured spectra as the one shown in figure~\ref{Fig:SPE} (left) and allow to quantify the resolution and the separation power between the noise pedestal and the single-photon response. In general, a better separation of the signal from the noise improves the energy threshold of a detector. Thus, an optimization of the two parameters as a function of gain is crucial for an optimal performance of XENON1T.

Figure~\ref{Fig:spe_quant} (left) shows the SPE resolution as a function of gain for the XENON1T PMTs, measured at room temperature. The yellow markers indicate the mean for binned values of the gain. It can be seen that the resolution of approximately 27\,\% does not improve significantly for gains above $3\times 10^6$. Figure~\ref{Fig:spe_quant} (right) shows the data for the peak-to-valley ratio. 
These results suggest an optimal operation of the PMTs at gains between $(2-3)\times 10^6$, given that low gains reduce signal saturation in the read-out electronics and low bias voltages minimize the stress on the PMTs during long-term operation.

\subsection{Transit time measurement} \label{sec:tt}
The timing performance of a PMT is characterized by the transit time spread (TTS), determined from the distribution of the time difference between the LED trigger (emission of the photon) and the arrival of the electron avalanche at the anode. This is relevant since a signal in XENON1T is generally observed by more than one PMT. Hence, the TTS defines the time window for coincident events between PMTs. A smaller time window allows to reduce possible random coincidences.
 
\begin{figure}[h!]
  \begin{center}
  \makebox[\textwidth][c]{
   \includegraphics[angle=0,width=0.52\textwidth]{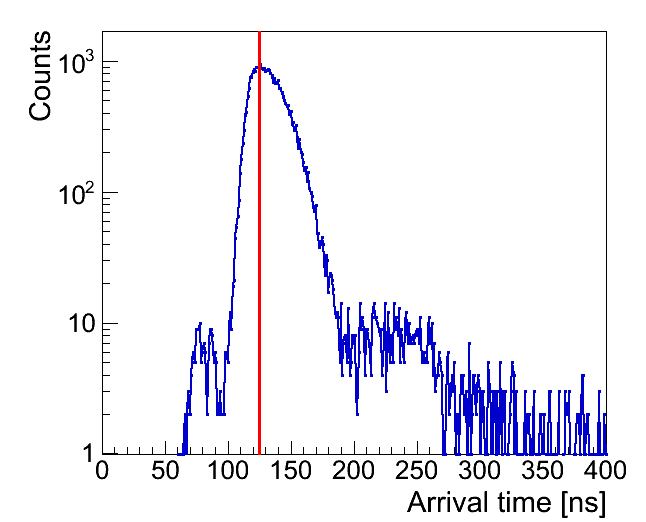}
\includegraphics[angle=0,width=0.52\textwidth]{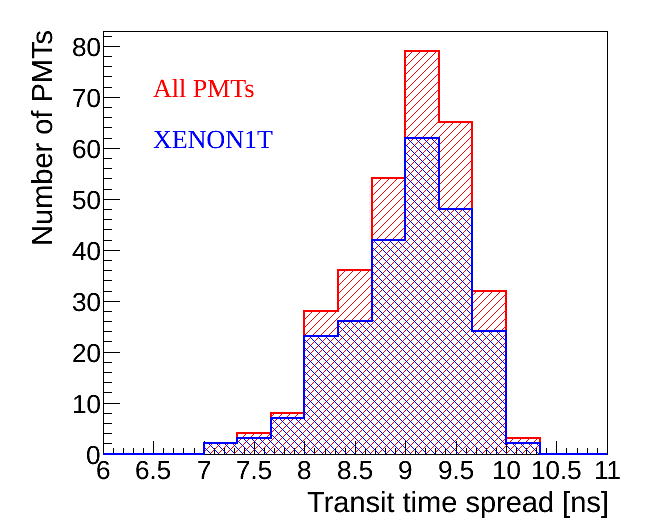}
}
 \end{center}	
  \caption[]{(Left) Transit time response for a single PMT channel with the most probable arrival time marked by a red line. (Right) Distribution of the transit time spread, defined as the FWHM of the main peak, for all (red) and selected (blue) tubes. \label{fig:TT} } 
\end{figure}

The TT has been measured with the TDC described in section\,\ref{sec:mpik_setups}. Figure~\ref{fig:TT} (left) shows the TT spectrum of a single PMT channel measured at room temperature, with the LED light set to SPE intensity in order to measure one signal per trigger in the TDC. In this LED mode, 10\,\% of the acquired signals are caused by single photoelectrons. The arrival of the electron avalanche at the anode is approximately 125\,ns after the LED trigger. This TT also includes the time delay from the electronics of the setup. The TT estimated by Hamamatsu for the photoelectron only, between the photocathode and the first dynode, is 46\,ns. The faster signals seen before the main peak are likely caused by photons passing through the photocathode and interacting directly with the first dynode. The events occurring after the main peak can be caused by photoelectrons backscattering off the first dynode~\cite{Kaether:2012bm}.
The TTS is defined as the FWHM of the main peak. This spread is a result of the different trajectories the photoelectrons take through the PMT according to their initial velocities and positions at the photocathode~\cite{Kaether:2012bm} as well as their emitted direction. The resulting TTS distribution for all PMTs is shown in figure~\ref{fig:TT}\,(right) with a mean of ($9.2 \pm 1.3$)\,ns. If this spread is convolved with the 1.4\,ns pulse width of the LED, the TTS for the tubes alone is reduced to ($9.1 \pm 1.3$)\,ns, showing that the LED pulse width has a negligible impact on the measurements. These results are in agreement with the TTS of 9\,ns reported by Hamamatsu.
\subsection{Dark count rates} \label{sec:dcrates}

The dark count (DC) rate of a PMT describes the number of signals per second above a given threshold in the absence of a light source. These signals are produced mainly by thermal electrons emitted when the PMT is powered at a certain voltage. At cryogenic temperatures, the thermal emission contribution is suppressed and a non-thermal DC remnant is present, which is caused, in part, by electron field emission (due to the high bias voltage), as well as internal and external radioactivity and cosmic particles interacting with the PMT. A low dark count rate in XENON1T is important to reduce accidental coincidences between PMTs. These random coincidences can imitate the detection of scintillation signals in the detector and, hence, contribute to the background during a dark matter search.

\begin{figure}[h!!]
  \begin{center}
  \makebox[\textwidth][c]{
   \includegraphics[angle=0,width=0.55\textwidth]{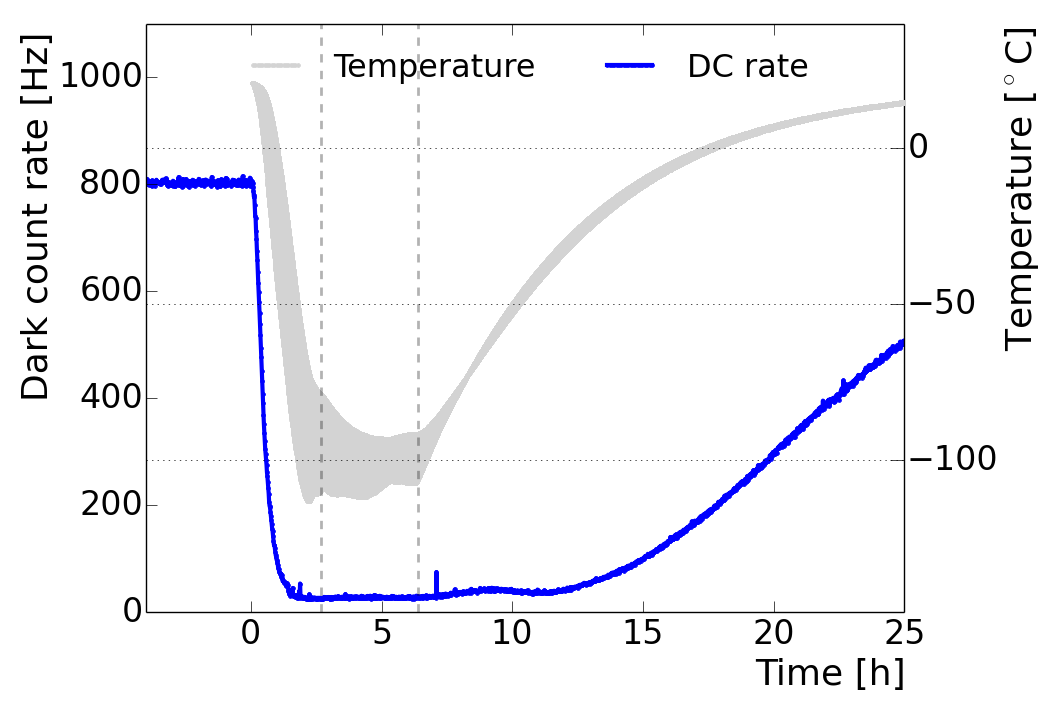}
 \includegraphics[angle=0,width=0.50\textwidth]{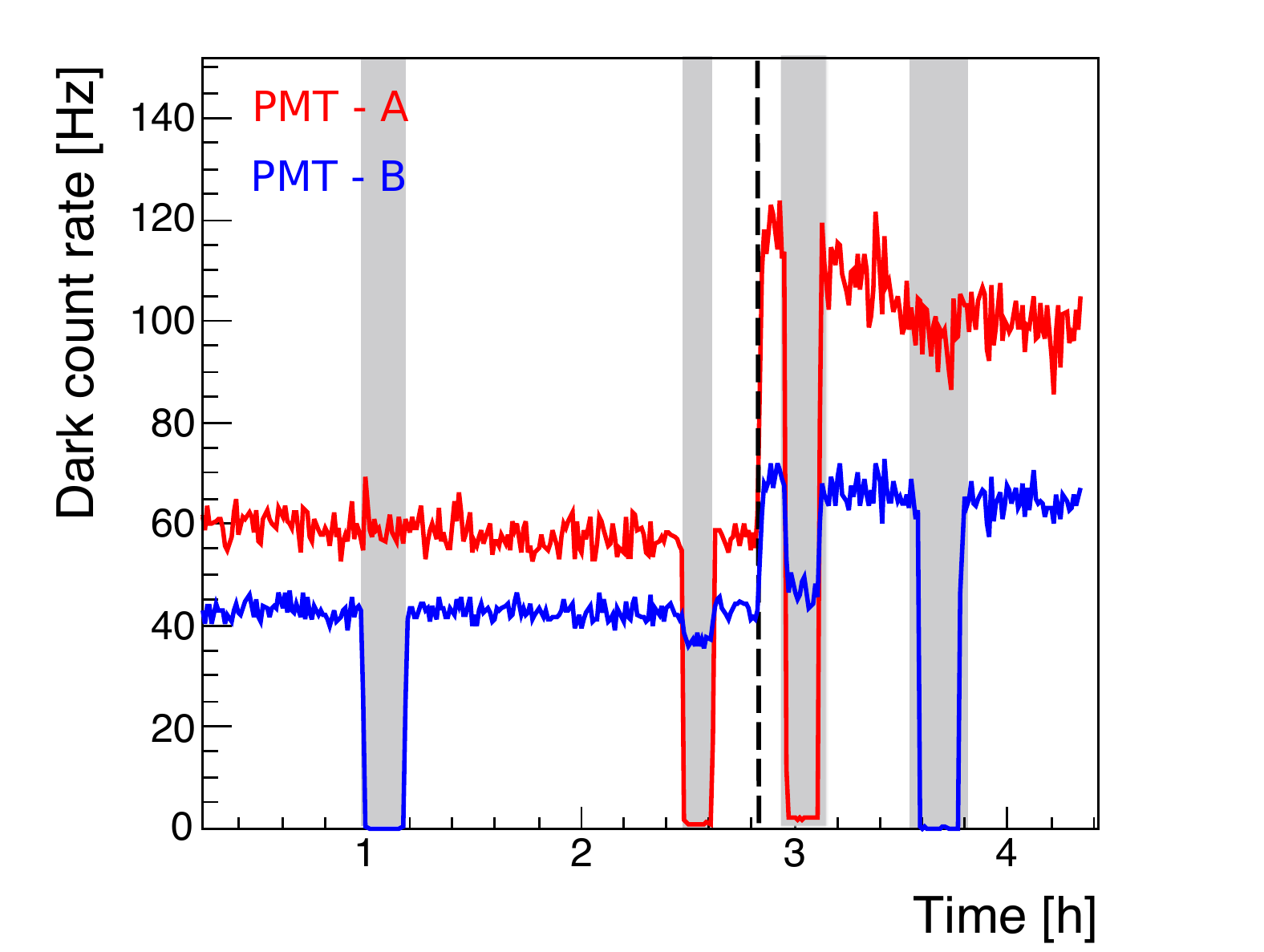}
 }
  \caption[]{(Left) Dark count rate evolution during cool down (blue curve) for a bias voltage of 1500\,V. The shaded grey area indicates the temperature range between the lower PMT array (position 3 in figure~\ref{Fig:setup}) and the upper PMTs (position 6). The vertical dashed lines indicate the time period of stable temperature. (Right) During stable temperature operation, tests for light emission are performed by measuring changes in the DC rates (see text). \label{Fig:DC_evl}}
  \end{center}	
\end{figure}

Figure~\ref{Fig:DC_evl} (left) shows an example of a DC rate measurement during a cooling cycle in the setup described in~\ref{sec:mpik_setups}. The time evolution of the dark count rate and its dependence on temperature is shown. The measurement starts at room temperature, with the DC rate dropping sharply at T = 0, indicating the beginning of the cool down at an average rate of 1.5\,K/min. Already at around $-30^{\circ}$C the DC rate reaches values close to its minimum. During the cooling process, the DC rate can show some instabilities, possibly due to small discharges inside the tube. After approximately 2.5 hours the temperature has decreased and stabilized at around $-100$\,$^{\circ}$C, where the DC rate remains at an approximately constant value. The performance of the tube is evaluated during the stable period until the cooling of the chamber is stopped. As the system heats up, the DC rate of the PMT also rises to its previous value at room temperature. The observed hysteresis of the DC rate during warming up is explained by the larger heat capacity of the tube with respect to the nitrogen vapour as measured by the thermometers. A minimum of 2 cool downs are performed for every PMT, during which different settings of the bias voltage are used: 1500\,V, 1680\,V and the voltage corresponding to a gain of $3\times10^6$.

Figure~\ref{Fig:DC_evl}\,(right) shows the dark count rates of two PMTs facing each other during the period of stable temperature at $-100$\,$^{\circ}$C. To the left of the dashed line, the PMTs are operated at a gain of $3\times10^6$, corresponding to voltages between $(1300-1500)$\,V. To the right of the line, the voltages have been increased to 1680\,V. The grey areas indicate time periods where one of the two PMTs was switched off to study its level of light emission. It can be seen that PMT-A emits light given that, when it is turned off, the opposite PMT shows a reduced dark count rate. Furthermore, the reduction is larger at 1680\,V ($\Delta$ = 25\,Hz) than at lower voltages ($\Delta$ = 10\,Hz), indicating a correlation with the bias voltage. A detailed study of this light emission effect is presented in section~\ref{sec:lightemission}.

\begin{figure}[h!]
  \begin{center}
  \makebox[\textwidth][c]{
   \includegraphics[angle=0,width=0.52\textwidth]{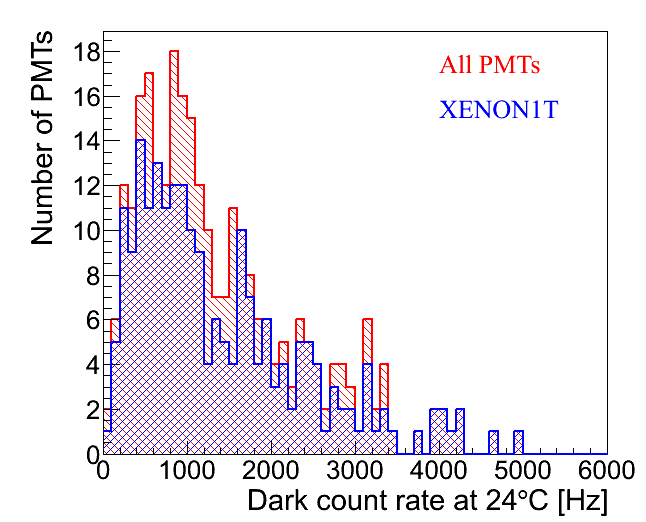}
   \includegraphics[angle=0,width=0.52\textwidth]{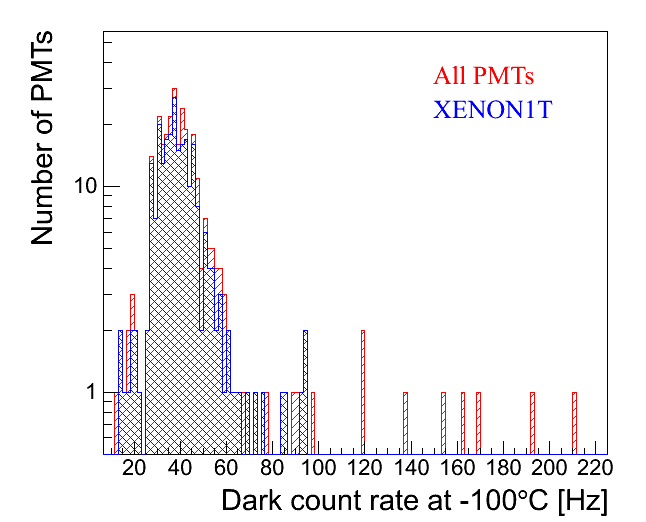} 
   }
   \caption[]{(Left) Distribution of the dark count rates at an equalized gain of $3\times 10^6$ at room temperature. (Right) Dark count rates at $-100$\,$^{\circ}$C with the same HV values as at room temperature. The red histograms show the data for all tested PMTs, while the blue histograms correspond to the PMTs selected for XENON1T.}
   \label{Fig:DC_quant}
  \end{center}	
\end{figure}

Figure~\ref{Fig:DC_quant} shows the histograms of the obtained dark count rates for all tested (red) and the XENON1T PMTs (blue) at an equalized gain of $3.0\times 10^6$ at room temperature (left) and at $-100$\,$^{\circ}$C (right).

The average dark count rate shows a value of 1348\,Hz (1400\,Hz) at room temperature for all (XENON1T) tubes with a large
standard deviation of around 988$\,$Hz (1025$\,$Hz). However, after cooling the tubes to $-100$\,$^{\circ}$C, the mean DC rate reduces to 44\,Hz (40\,Hz) with a smaller standard deviation of 24\,Hz (13\,Hz) for all 321 tested (selected) tubes. In the right figure, the red histogram shows a population of PMTs with rates above 100\,Hz which have been rejected due to light emission or an unstable dark count rate. It is important to mention that the measured dark count rates depend on the pulse height threshold at which the counter trigger is set. For these measurements a threshold at $\sim$\,$\frac{1}{4}$\,PE has been used, determined at room temperature. As shown in section~\ref{sec:xetests}, the measured gain at $-100$\,$^{\circ}$C changes with respect to the values at room temperature. An average gain increase of 10\,\% has been measured in this setup, with which the counter threshold corresponds to 0.15\,PE and increases the single PE acceptance by about 2\,\%. The rates also depend on the shielding of the test setup and the distance between the opposing PMT structures. A larger distance between the PMT arrays increases the measured values and standard deviation of the DC rates, likely due to an increased solid angle to observe Cherenkov light or scintillation (390\,nm wavelength photons~\cite{Lehaut:2015cez}) in the N$_2$ gas from ionizing radiation such as cosmic rays and $\alpha$ particles. For the presented results the distance between two PMT windows is 2\,cm.

\section{Detection of light emitting PMTs} \label{sec:lightemission}
During the DC rate measurements described in section~\ref{sec:dcrates}, some cases of light emission from the phototubes have been identified. This behaviour is problematic for rare event searches, such as dark matter detection, since it may produce coincident pulses in the detector that can be interpreted as a scintillation signal. If the level of light emission is too large, those accidental coincidences might create an irreducible background to the dark matter search and could significantly limit the sensitivity of the detector. The quantification of such light emission is not straightforward since it depends strongly on the setup and ambient conditions. However, a qualitative study has been made by observing a change in the dark count rates of either the light emitting PMT itself or of the PMT facing it. The absolute change of the dark count rate will depend on the amount of collected light, which, in turn, depends on the solid angle between the opposing tubes. This method is thus not an absolute measurement of the total light emission, and the quantitative results may vary in XENON1T.

Two main categories of light emission have been observed in the extensive testing campaign described here. In some cases, the tube emits a large amount of light within a period of a few seconds. This so-called ``flash'' is observed by the tube itself and the  PMTs operated on the opposite array, increasing the measured rates. The rates can take from several minutes up to a few hours to decrease to the previous levels, possibly due to the exposure of the PMT window to the light from this event. The microscopic processes causing this spontaneous and strong light emission are not completely understood, but, given their rare occurrence (in only $\sim$\,1\,\% of all PMTs tested) and straight-forward identification, tubes in which only one such event has been observed were not rejected.

\begin{figure}[h!]
\begin{center}
\makebox[\textwidth][c]{
\includegraphics[angle=0,width=0.52\textwidth]{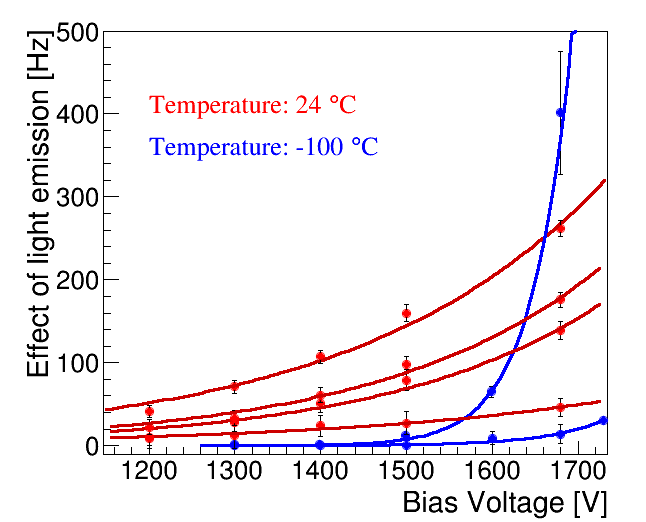}
\includegraphics[angle=0,width=0.52\textwidth]{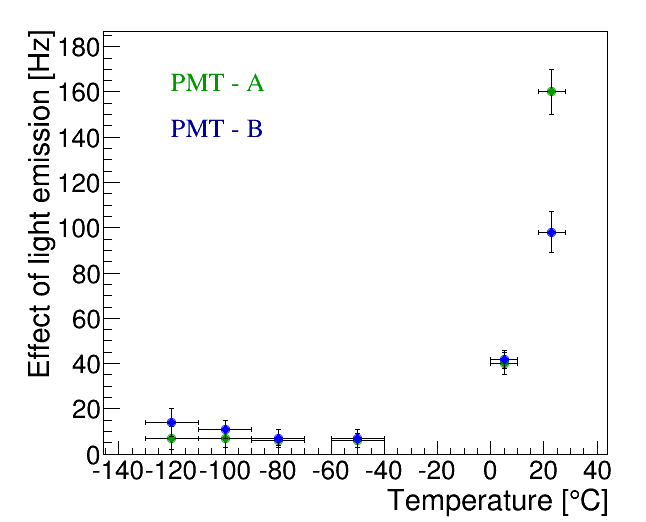}
}
\caption[]{(Left) Change of the dark count rate due to a light emitting PMT with respect to the applied bias voltage of the light emitting PMT. The HV of the monitoring PMT is kept constant to 1500\,V during the scan. It can be seen that the amount of emitted light is dependent on the HV, and if the tube is operated at sufficient low bias voltages, light emission is not observable with this method. The effect of light emission for different PMTs at room temperature is indicated by the red lines, while the measurements at $-100\,^{\circ}$C are shown in blue. (Right) Correlation between the effect of two light emitting PMTs at different operation temperatures and a constant bias voltage of 1500\,V.}
 \label{fig:hv_le}
\end{center}
\end{figure}

The second type, denominated ``micro light emission'', is harder to observe since it is only detectable by a small change of the dark count rate of the opposing PMT, as explained previously in figure~\ref{Fig:DC_evl}. The level of micro light emission is defined by the difference of the average dark count rate induced by the tested PMT which is turned on and off. This method allows to quantify the observed light emission for several HV values of the light emitting PMT and the functional dependence on the HV can be derived, as shown in figure~\ref{fig:hv_le} (left). Results are presented for measurements at room temperature and at $-100\,^{\circ}$C for several PMTs. 

The effect of light emission decreases with lower bias voltages and is larger at higher temperatures while it is typically suppressed at cryogenic temperatures. This method, however, might not be sensitive enough to detect a small level of micro light emission at low bias voltages. Figure~\ref{fig:hv_le} right shows the temperature dependence of the effect at 1500\,V. 
Light emission has been measured in a total of 53 PMTs. These have been returned to Hamamatsu and replaced. Other studies of the R11410 PMT have also shown light emission due to processes within the tube~\cite{Akimov:2015cta}\cite{Akimov:2015alu}. The effects of light emission of this PMT type operated inside a dark matter experiment are shown in~\cite{Li:2015qhq}.


\section{Measurements in gaseous and liquid xenon} \label{sec:xetests}

After the PMTs have been tested at room temperature and cooled down in a nitrogen environment, a subset has been selected for testing in both gaseous and liquid xenon. This selection includes PMTs with unstable behaviour during the general testing described earlier or hints of possible leaks, as will be discussed in section~\ref{sec:ap_leaks}.

\begin{figure}[h]
  \begin{center}
   \includegraphics[width=0.85\textwidth]{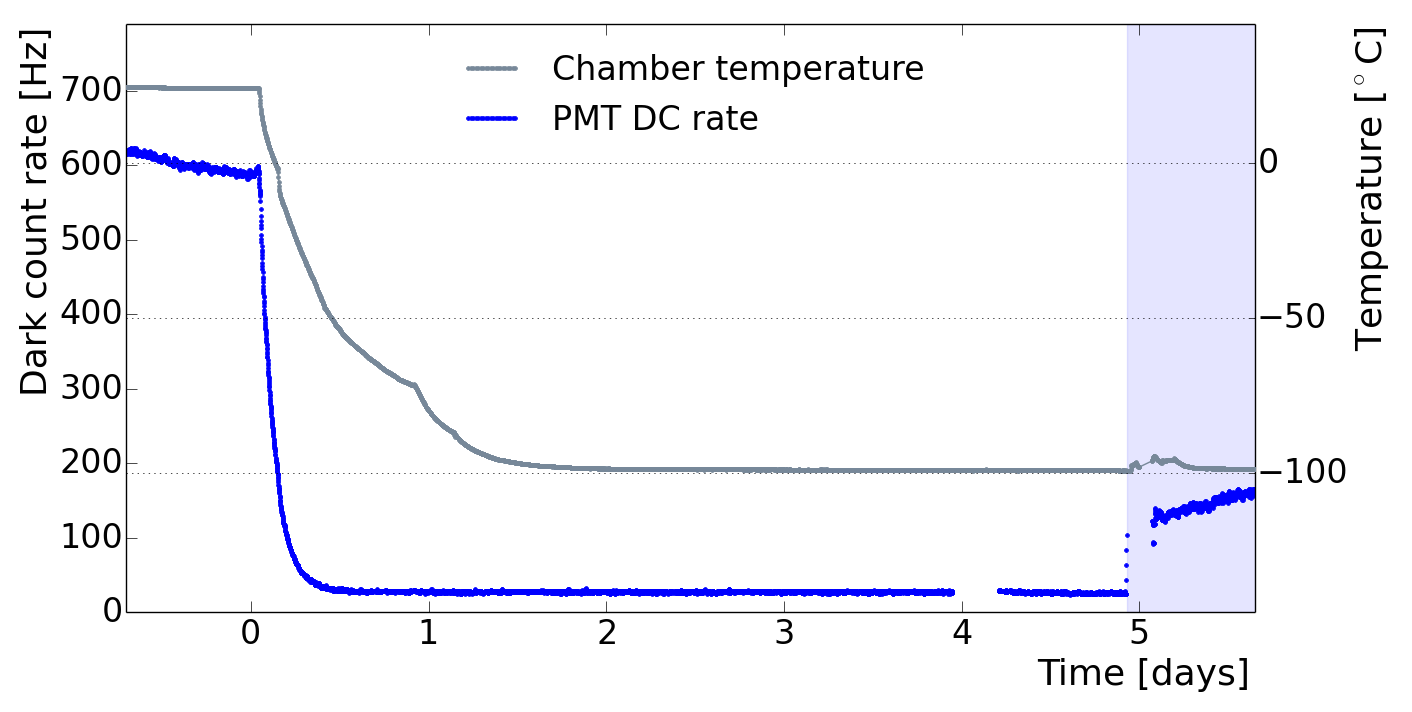}
   \caption[]{Evolution of the dark count rate of a PMT during operation in gaseous and liquid xenon. The gray line indicates the temperature of the inner chamber during cool down. The rate (blue line) decreases from hundreds of Hz at room temperature to an average of 28\,Hz when operated in the gas phase, at around $-100\,^{\circ}$C, remaining stable within 2\,Hz. The blue shaded area indicates a filling of liquid xenon, after which the rate increases to about 200\,Hz due to the scintillation light in the liquid phase.}
  \label{Fig:MXL_cool_down}
  \end{center}	
\end{figure}

As described in section~\ref{sec:mxl_setup}, the xenon testing facility allows 5 tubes to be operated simultaneously. The measurements are performed over a period ranging from a couple of weeks to several months. The cooling of the system is achieved by filling gaseous xenon (GXe) into the chamber over several steps. In the first step, after reaching a vacuum pressure below 2$\times 10^{-5}$\,mbar, around 0.1\,kg of xenon gas is introduced into the chamber, raising the absolute pressure to 1.8\,bar. As the PTR cools the system and the pressure decreases close to 1.4\,bar, more xenon is filled to increase the cooling rate. The set point around $-100\,^{\circ}$C is reached in approximately 24~hours, corresponding to an average cooling rate below 0.1\,$^{\circ}$C/min. For measurements in liquid, 2.8\,kg of xenon are filled to cover the PMT body. Figure~\ref{Fig:MXL_cool_down} shows an example of the cooling cycle for a PMT in xenon. Several such cool downs are performed for each PMT to test the stability and resistance to thermal cycling.

\subsection{Dark count rate stability}

The xenon testing facility allows for long-term measurements of PMT properties given that it operates at very stable conditions for long periods of time. During cryogenic operation, the temperature in the cryostat can be kept stable within 0.2\,$^{\circ}$C during a day. The changes over a month of operation are no larger than 2\,$^{\circ}$C. The pressure, in turn, varies within 0.03\,bar in a day and up to 0.07\,bar within a month.

The blue curve in figure~\ref{Fig:MXL_cool_down} shows an example of the dark count rate of an R11410-21 PMT at 1500\,V during cool down. The average dark count rate of 30 PMTs tested at $-100\,^{\circ}$C in GXe was measured to be ($46\pm4$)\,Hz over several weeks. This represents a stability of around 10\,\% for the dark count rate of the XENON1T PMTs at cryogenic temperatures. The average DC rate in cold GXe is in agreement with the cold test results from section~\ref{sec:dcrates}, in which the mean rate was ($44\pm23$)\,Hz. When filling the chamber with LXe, the PMT dark count rate increases from tens of hertz to a few hundred hertz. This is due to scintillation light produced by interactions of cosmic and ambient radiation in the liquid medium.

\subsection{Gain evolution}\label{sec:gain}

\begin{figure}[]
  \begin{center}
   \includegraphics[width=0.99\textwidth]{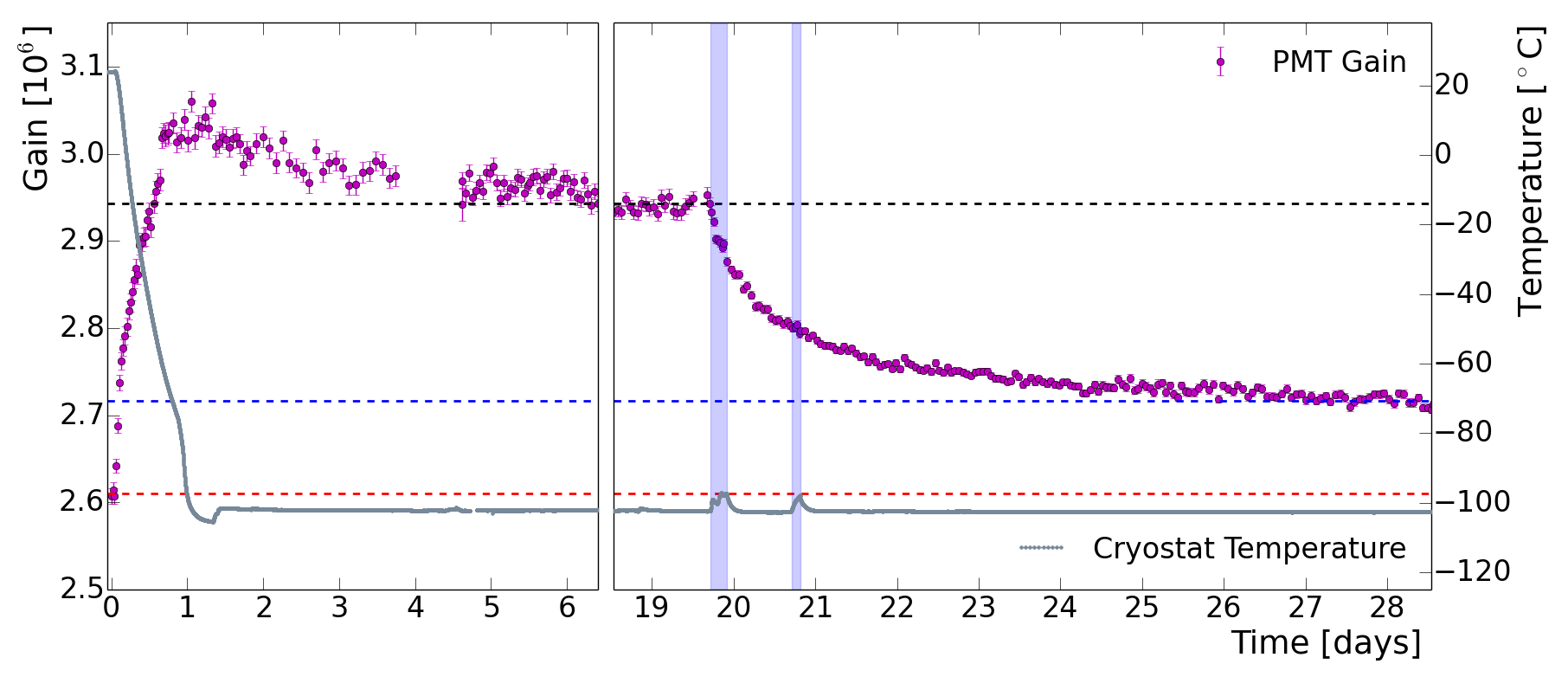}
   \caption[]{PMT gain evolution during cool down in gas xenon (first 20 days) and liquid xenon (filling on days 20 and 21, indicated by the blue shaded areas). An increase of around 12\,\% in the gain is measured between room temperature and $-100\,^{\circ}$C gaseous xenon. After the filling of liquid xenon, a decrease of around 7\,\% is observed in the gain with respect to the value in the gas phase.}
   \label{Fig:gain_evolution}
  \end{center}	
\end{figure}

The gain of the PMTs has been measured at cryogenic temperatures both in gas and liquid xenon. Figure~\ref{Fig:gain_evolution} shows the gain evolution of a PMT during cool down from room temperature to $-100\,^{\circ}$C in a xenon environment.

When cooled and maintained in GXe, the gain increases and settles at a value higher than at room temperature. This effect is explained by the decrease in electric resistance of the signal cables within the cryogenic setup, which results in a higher charge output. The effect has been estimated for the nitrogen cooling setup at MPIK, yielding the expected results for the decrease in the resistivity of the cables. On the other hand, when covered in LXe, a decrease of the PMT gain is observed. The direct contact of the PMT body with the LXe (window first) accelerates the thermalization of its inner components (dynodes), reducing their temperature more efficiently. The temperature dependence of the dynode secondary emission can thus play a role in the decrease of the PMT gain. The same observation has been reported in~\cite{Baudis:2013xva}.

For the PMTs tested in xenon, the average increase of the gain between room temperature and GXe at $-100\,^{\circ}$C has been measured to be around 7\,\%, while an average decrease of 3\,\% is observed during operation in LXe with respect to the values in cold GXe.

 \subsection{PMT response to Xe scintillation light}
 
\begin{figure}[h]
\begin{center}
\makebox[\textwidth][c]{
\includegraphics[angle=0,width=0.5\textwidth]{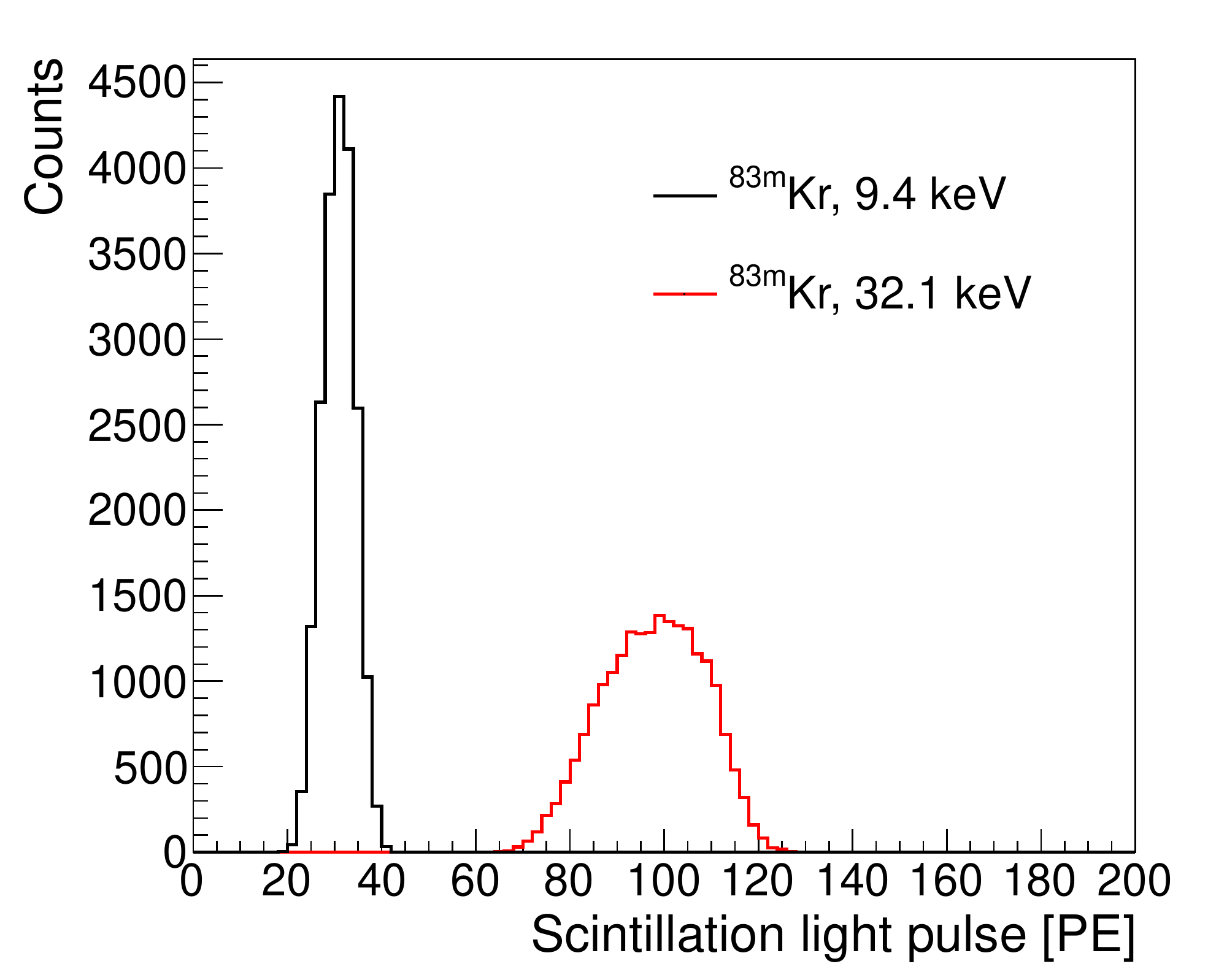}
\includegraphics[angle=0,width=0.5\textwidth]{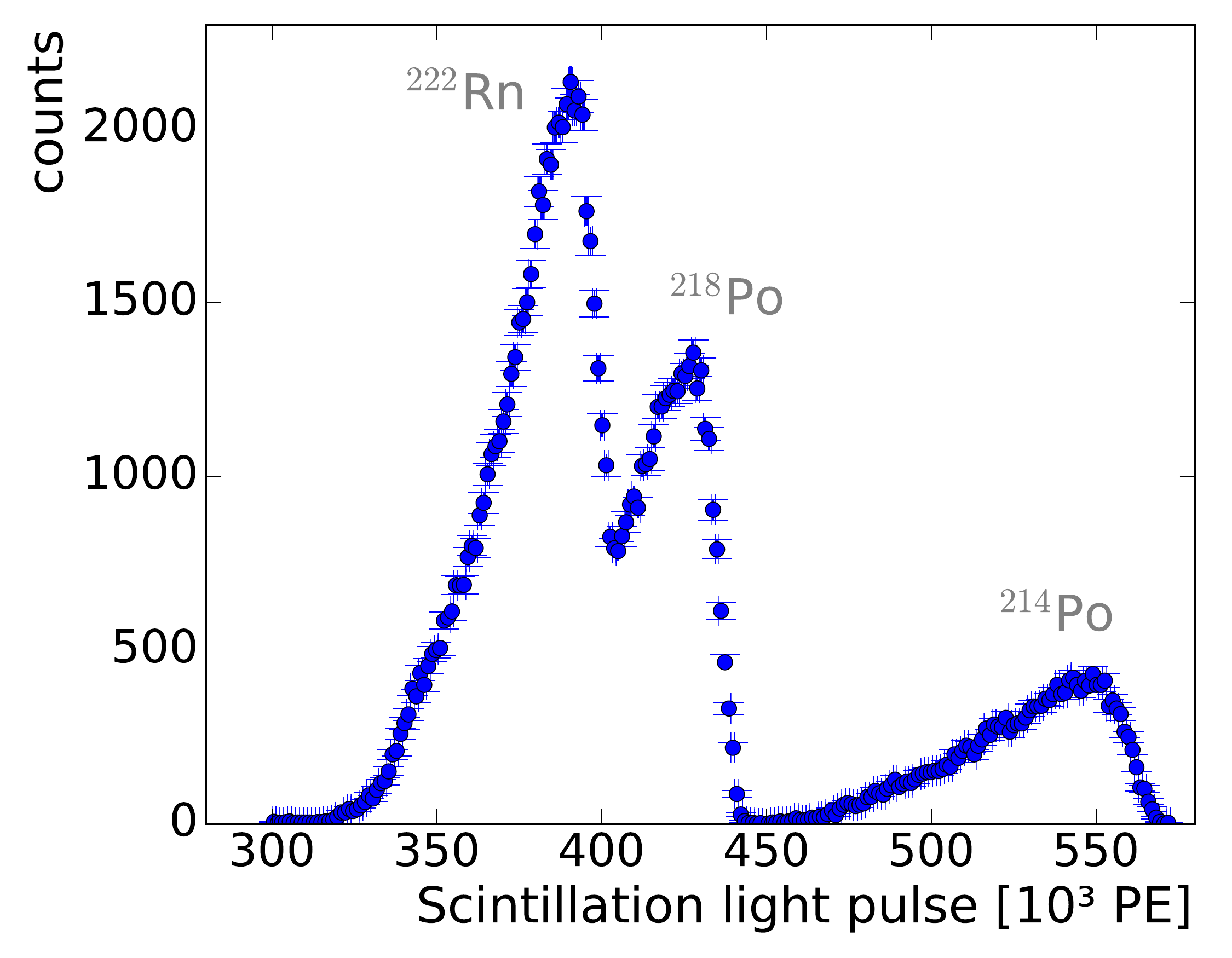}
}
\caption[]{(Left) Measurement of the Xe scintillation light from the two-step $^{83m}$Kr decay. The resolved pulses from the 32.1\,keV line (first signal) are shown in red, while the pulses from the 9.4\,keV line (second signal) are shown in black. (Right) Measurement of $^{222}$Rn decay and its daughters measured in a LXe chamber~\cite{Bruenner:2016ziq}.}
 \label{fig:cal_peak}
\end{center}
\end{figure}

The PMT response to xenon scintillation light has been studied. For this, the tubes were exposed to two different calibration sources while operated in single phase detectors (LXe at $-100\,^{\circ}$C). First, a $^{83m}$Kr source has been injected into the xenon testing facility described in section~\ref{sec:mxl_setup} with one PMT in operation. The source is produced by the decay of $^{83}$Rb via pure electron capture. The $^{83m}$Kr isomeric state, located 41.5\,keV above the ground state, decays with a half-life of 1.83\,h to the $^{83}$Kr excited state at 9.4\,keV, producing a first signal. The subsequent decay to the ground state, with a half-life of 154\,ns, produces a second signal. A detailed discussion can be found under reference~\cite{Manalaysay:2009yq}. Figure~\ref{fig:cal_peak} (left) shows the measured scintillation signal from the two energy lines of $^{83}$Kr decay with the R11410-21 PMT.

The events from $^{83}$Kr have been used to calculate the distribution of delay time between the first and second signals. A fit to the distribution results in a half-life of $(155.4 \pm 1.3)$\,ns after basic data quality cuts. The derived half-life is consistent with the published value of $(154.4 \pm 0.1)$\,ns~\cite{nuctable}. This agreement confirms the detection of $^{83m}$Kr decays.

Additionally, two PMTs have been tested in a LXe chamber described in~\cite{Bruenner:2016ziq}. The PMT windows face each other and are operated in coincidence to efficiently reduce background. Enriching the LXe with radon by employing an aqueous $^{226}$Ra source, allows to measure decays of $^{222}$Rn and its daughters. The spectrum of the measured events from interactions of alpha particles can be seen in figure~\ref{fig:cal_peak} (right) where the peaks represent the detection of $^{222}$Rn (5.59\,MeV), $^{218}$Po (6.11\,MeV) and $^{214}$Po  (7.83\,MeV) $\alpha$-decays\,\cite{nuctable}. In addition, we identify $^{214}$Po $\alpha$-decays which occur directly after the $\beta$-decay of their $^{214}$Bi mother nuclide due to the short lifetime of $^{214}$Po with T$_{1/2} = (162.3 \pm 1.2)$\,$\mu$s~\cite{nuctable}. From the data acquired with this setup, a value of
 T$_{1/2} = (145 \pm 2_{\rm{stat.}} \pm 10_{\rm{sys.}})$\,$\mu$s has been determined~\cite{Dominick:thesis} and both values agree within the uncertainties.

These results, together with the measurement of the $^{57}$Co spectrum in~\cite{Baudis:2013xva} verify that the R11410-21 PMTs are sensitive to the xenon scintillation light, as has also been shown in~\cite{Cao:2014jsa}\,\cite{Akimov:2012aya}\,\cite{Lyashenko:2014rda}.

\section{Study of afterpulses and leak identification} \label{sec:ap}

A detailed study of the afterpulses in the R11410-21 phototube has been performed following the methods introduced in~\cite{Baudis:2013xva}. The results presented here focus on the evolution of the afterpulse rates and spectra after operation of the phototubes in cold Xe. The methods for the identification of ions within a PMT are detailed and implemented for the diagnosis of leaks which can appear when cooling down the tubes. 

\subsection{Ion identification through afterpulse timing} \label{sec:ionid}

Residual gas molecules in the PMT vacuum comprise one of the sources for afterpulsing. In its trajectory from the photocathode to the first dynode, a photoelectron may ionize a molecule, which in turn drifts to the photocathode and generates an afterpulse. A specific ion can be identified by the time delay between the afterpulse and the main pulse that generated it. An ion generated at the position $s_0$ between the focusing grid (located at a distance L) and the PMT photocathode (located at 0) will have the following travel time $t$:

\begin{equation} \label{eq:tintegral1}
t = \int^0_{s_0} \frac{1}{\rm{v(s)}} ds = \sqrt{\frac{m}{2q}} \int^0_{s_0} [V(s_0)-V(s)]^{-1/2} ds ,
\end{equation}

\noindent where the velocity \textit{v(s)} as a function of position has been determined with the Lorentz force equation in an electric field. The mass and charge of the ion are \textit{m} and \textit{q}, respectively, and $V(s)$ is the electric potential as a function of position.

At first approximation, the field between the photocathode and focusing grid may be considered that of a parallel plate capacitor. However, for PMTs such as the 3-inch R11410 models, the field lines converge towards the focusing grid, for which a better approximation is obtained with a quadratic potential of the form:
$V(s) = V_0 \left(\frac{s}{L} \right)^2$ (see also~\cite{Ma:2009aw}). The introduction of this potential in equation~\ref{eq:tintegral1} yields the following result:

\begin{equation} \label{eq:aptiming1}
t = \sqrt{\frac{m}{2q}} \left.\frac{L}{\sqrt{V_0}} \textrm{arcsin}\left(\frac{s}{s_0}\right)  \right|^0_{s_0}
 = \frac{\pi}{4} \sqrt{\frac{2m}{qV_0}} L,
\end{equation}

\noindent which is independent of the position of ionization. This equation can be further simplified by introducing the proton mass and charge, substituting 1C~=~(kg\,m$^2$)/(V\,s$^2$) and rewriting the terms in units of cm and $\mu$s:

\begin{equation} \label{eq:aptiming3}
t = \left( 1.134 \, \frac{\textrm{V}^{1/2} \mu \textrm{s}}{\textrm{cm}} \right)  \sqrt{\frac{L^2}{V_0} \frac{M}{Q}}, 
\end{equation}

\noindent with $V_0$ in volts, $L$ in cm and $t$ in $\mu s$. M/Q is the mass-to-charge ratio of the ion, set in number-of-nucleons/units-of-charge and is thus dimensionless. In the case of the R11410-21 tube, L~=~4.1\,cm. When operated at 1500\,V, using the XENON1T voltage divider, the potential difference corresponds to  $V_0$~=~323.4\,V. Table~\ref{table:aptimes} shows the identified ions from the afterpulse analysis, comparing the experimental measured times, the calculated times from equation~\ref{eq:aptiming3} and the times obtained from the simulations described in the following paragraph. The energy of the electrons reaching the focusing grid, given the potential difference previously stated, is around 324\,eV. This energy is sufficient to ionize these molecules, since their ionization potential is in the order of tens of eV.  

\begin{table}
\begin{centering}
\begin{tabular}{c c c c c}
\hline \multirow{2}{*}{Ion} & \multirow{2}{*}{M/Q} & \multicolumn{3}{c}{Afterpulse Time [$\mu$s]} \\
\cline{3-5}
& & Measured & Calculated & Simulated \\ \hline
H$^+$    &  1   & 0.28 $\pm$ 0.02 & 0.26 $\pm$ 0.01 & 0.27 $\pm$ 0.01\\ 
H$_2^+$  &  2   & 0.39 $\pm$ 0.02 & 0.37 $\pm$ 0.01 & 0.38 $\pm$ 0.01\\ 
He$^+$   &  4   & 0.52 $\pm$ 0.02 & 0.52 $\pm$ 0.01 & 0.52 $\pm$ 0.01\\ 
CH$_4^+$ &  16  & 1.01 $\pm$ 0.03 & 1.03 $\pm$ 0.02 & 1.02 $\pm$ 0.02\\ 
Ne$^+$   &  20  & 1.13 $\pm$ 0.03 & 1.16 $\pm$ 0.03 & 1.14 $\pm$ 0.02\\ 
N$_2^+$  &  28  & 1.33 $\pm$ 0.03 & 1.37 $\pm$ 0.03 & 1.33 $\pm$ 0.03\\ 
Ar$^+$   &  40  & 1.58 $\pm$ 0.04 & 1.63 $\pm$ 0.04 & 1.57 $\pm$ 0.03\\ 
Xe$^{++}$&  65  & 2.02 $\pm$ 0.05 & 2.08 $\pm$ 0.05 & 2.02 $\pm$ 0.04\\ 
Xe$^+$   &  131 & 2.85 $\pm$ 0.07 & 2.96 $\pm$ 0.07 & 2.80 $\pm$ 0.06\\
\hline 
\end{tabular}
\caption{Several residual molecules in the PMT have been identified as generators of afterpulses. M/Q corresponds to the mass-to-charge ratio in nucleons over units of charge. The time delay measured by experiment at 1500\,V is compared to the calculated value from equation~\ref{eq:aptiming3} and the result from simulations. The experimental errors correspond to the standard deviation of each peak. The calculated errors are propagated from the measurement uncertainty of the distance between photocathode and focusing grid, as well as the value of the voltage divider resistors. To quantify the uncertainty in the simulations, the grid-to-dynode distance and resistor values have been varied, and the distribution spread for each simulated ion has been taken into account.}
 \label{table:aptimes} 
\end{centering}
\end{table}

\begin{figure}[h]
  \begin{center}
   \makebox[\textwidth][c]{
   \includegraphics[angle=0,width=0.53\textwidth]{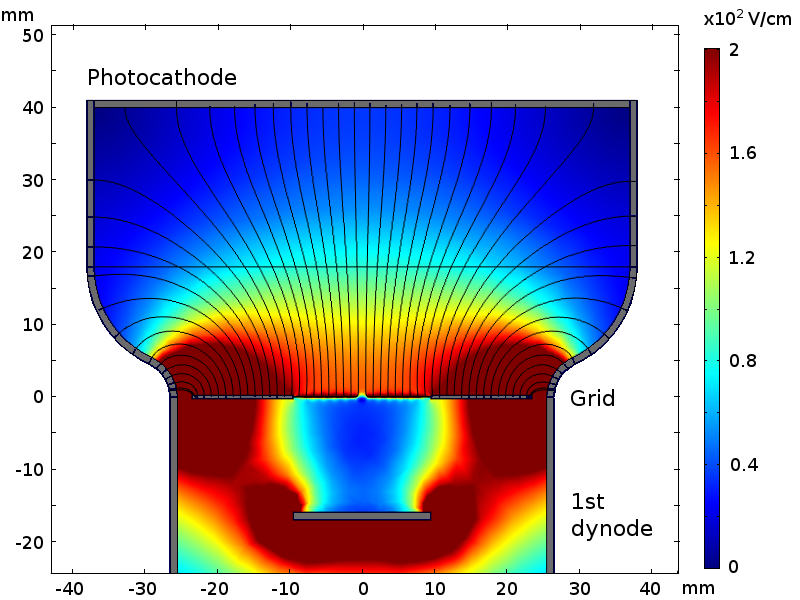}
   \includegraphics[angle=0,width=0.51\textwidth]{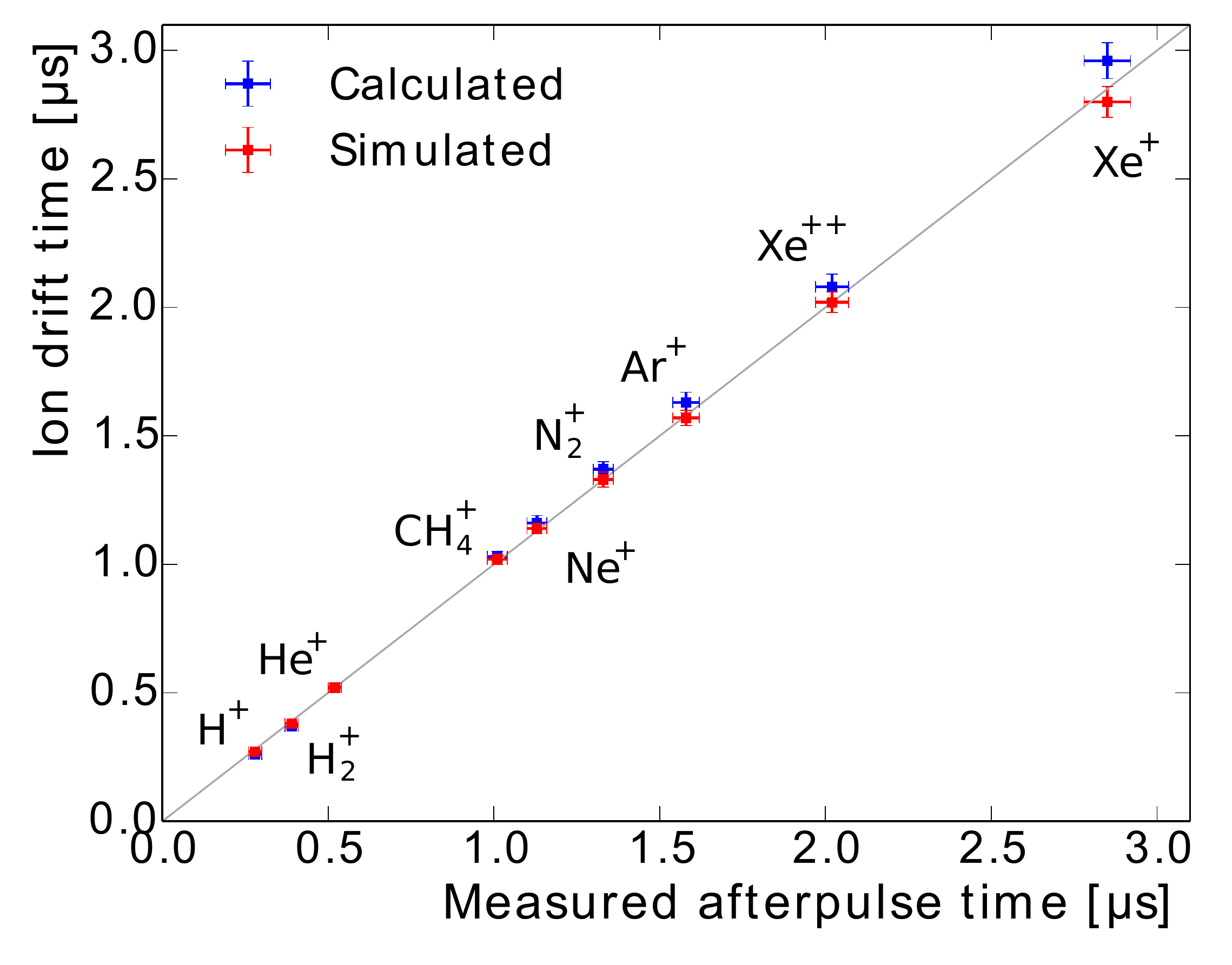}
   }
   \caption[]{(Left) Simulation of the PMT electric field using COMSOL Multiphysics\textsuperscript{\textregistered}. The color scale indicates the field strength in V/cm. Electron and ion transport have been simulated in the generated field, allowing to estimate their transit times. (Right) Comparison of the calculated (blue) and simulated (red) ion drift times with respect to the experimental afterpulse results. Values taken from table~\ref{table:aptimes}. The diagonal line indicates a one-to-one relation, for reference.}
   \label{Fig:Sim}
  \end{center}	
\end{figure}

For simulations of the electric field and transport of electrons and ions, a model of the PMT has been developed using the COMSOL Multiphysics\textsuperscript{\textregistered} software~\cite{COMSOL}. The dimensions have been taken from direct measurement of the elements in an opened tube. In particular, the focusing grid and dynodes have been reproduced in detail. Figure~\ref{Fig:Sim}\,(left) shows the simulation of the electric field in the PMT. The drift of electrons towards the first dynode has been simulated from a homogeneous distribution of electrons at rest on the photocathode surface. The average transit time was measured to be ($45.4\,\pm\,1.8$)\,ns, with a collection efficiency of 95\,\%. These values are in agreement with those reported by Hamamatsu (see section~\ref{sec:tt}), where the transit time is ($46\,\pm\,9$)\,ns and the photoelectron collection efficiency is around 90\,\%~\cite{Lung:2012pi}. Several of the ions experimentally identified to cause afterpulses have also been simulated. With an initial position at rest and distributed within the volume between the focusing grid and the photocathode, their corresponding drift times through the simulated electric field have been recorded. The results are presented in table~\ref{table:aptimes}. The experimental, calculated and simulated results differ on average by about 3\,\%, being in good agreement when taking into account the uncertainties described in the table caption. This suggests a high accuracy for ion identification. The comparison is also shown in figure~\ref{Fig:Sim}\,(right). The slight deviations between the calculated drift times for Xe ions with respect to the simulated and experimental values indicate that the quadratic potential implemented in section~\ref{sec:ionid} is a good approximation but not an exact description of the complex electric field in the PMT, shown in figure~\ref{Fig:Sim}\,(left).

\subsection{Afterpulse analysis and leak diagnosis} \label{sec:ap_leaks}

\begin{figure}[h]
  \begin{center}
  \makebox[\textwidth][c]{
   \includegraphics[angle=0,width=0.53\textwidth]{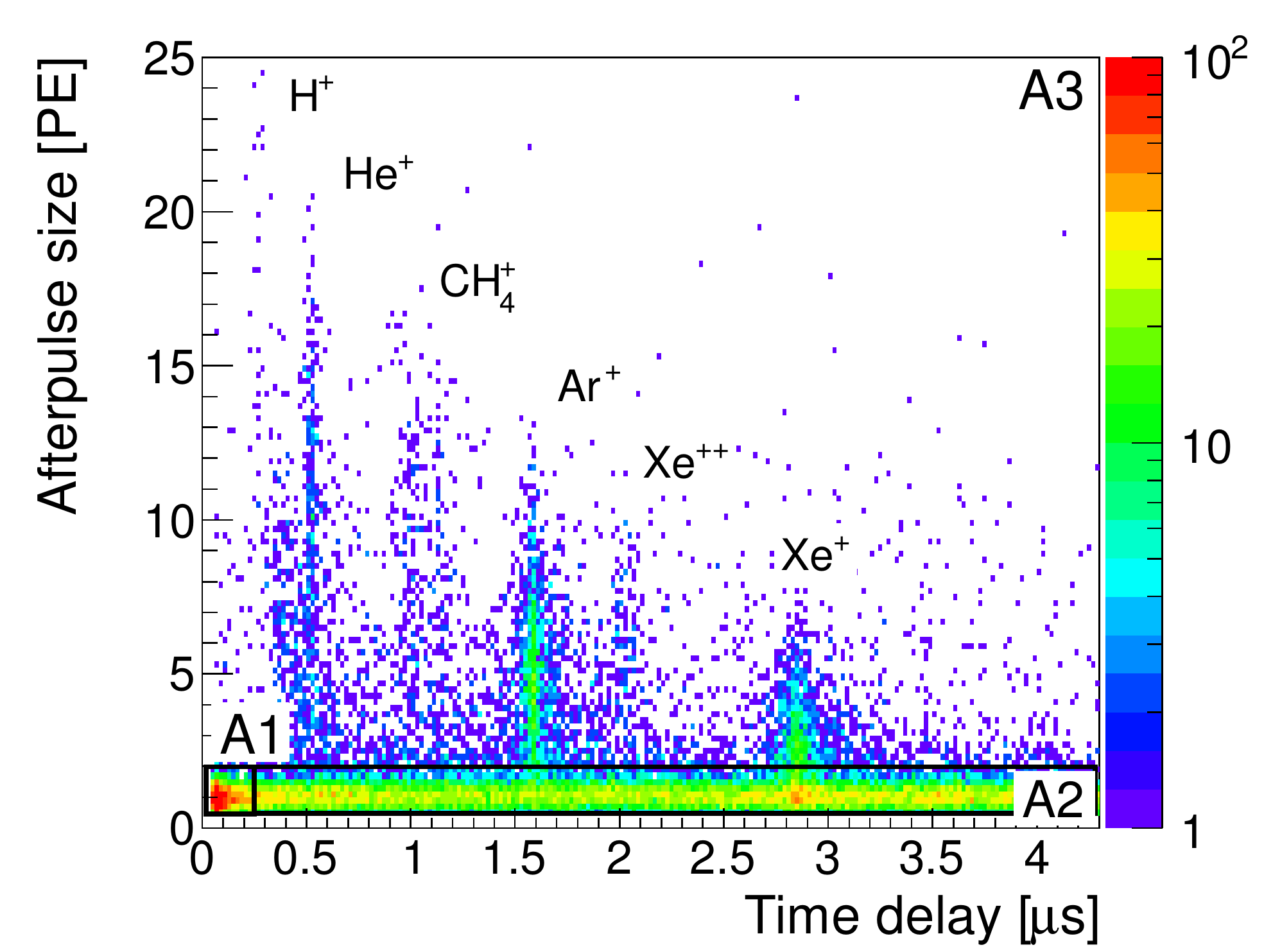}
    \includegraphics[angle=0,width=0.495\textwidth]{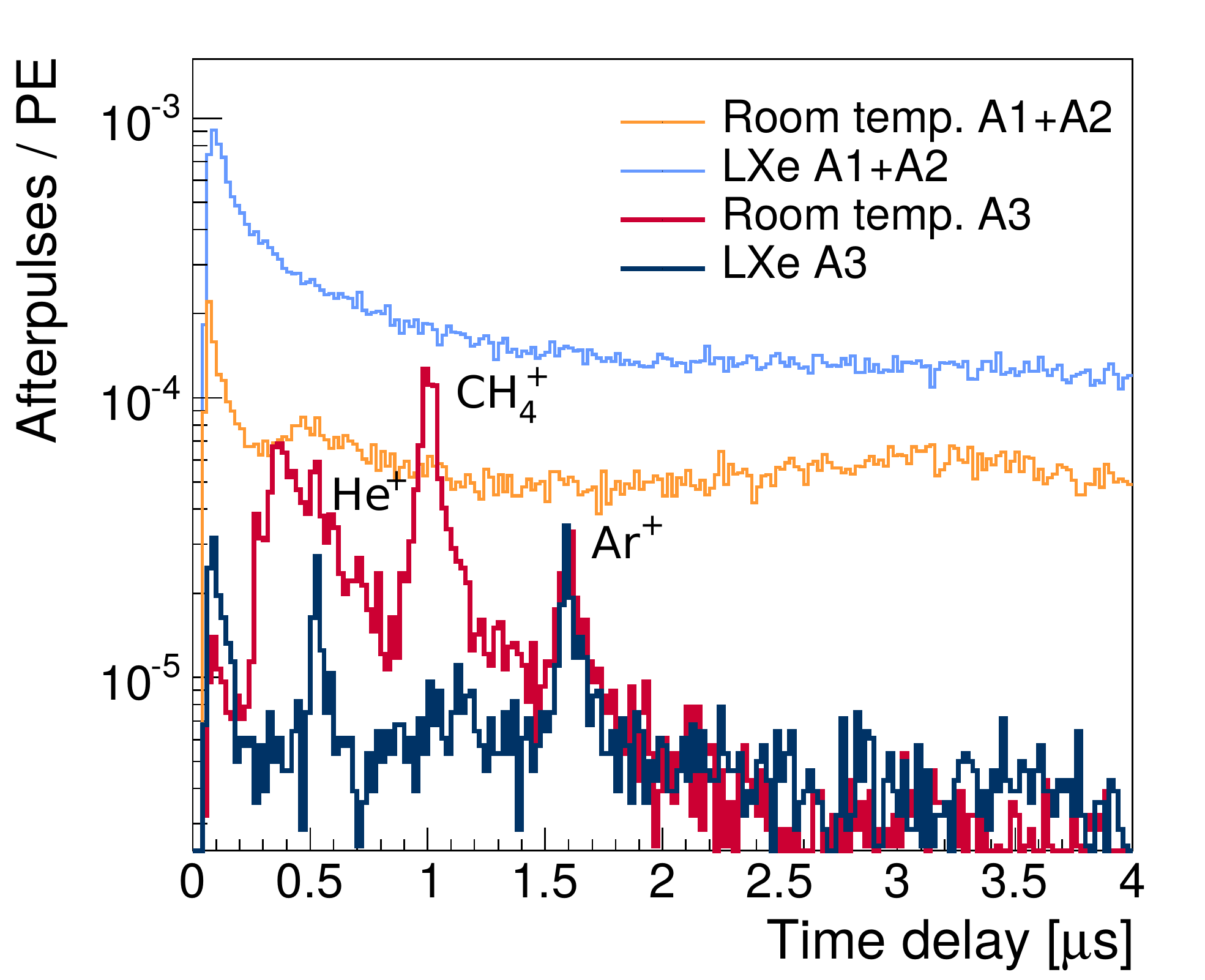}
    }
   \caption[]{(Left) Afterpulse size and time delay for a leaky PMT at room temperature after operation in LXe. The afterpulses are classified into groups A1, A2 and A3, according to their origin. The group A3 corresponds to afterpulses generated by ions of residual gas molecules in the PMT vacuum, which have been identified according to the analysis described in section~\ref{sec:ionid}. (Right) Comparison of afterpulses in A1+A2 and in A3 for a PMT at room temperature and in LXe.}
   \label{Fig:Afterpulses_mxl}
  \end{center}	
\end{figure}

The afterpulses generated by ions are characterized by amplitudes of several photoelectrons and appearance at specific times, determined by the mass and charge of the ion, as discussed before. Such afterpulses have been classified in the group A3 of figure~\ref{Fig:Afterpulses_mxl} (left). The figure shows the size of the afterpulses for a leaky PMT, normalized to that of a single photoelectron on the \textit{y} axis, while the \textit{x} axis corresponds to the afterpulse transit time. Other groups of afterpulses have also been identified according to their characteristics and origin: A1-- Pulses with a very short time delay (several tens of nanoseconds) and amplitudes around 1~PE, possibly generated by secondary electrons scattering off the first dynode which backtrack and successively strike again the first dynode. A2-- Pulses with a time delay of up to several microseconds and amplitudes around 1~PE. This population is generated by dark pulses and single electrons of unclear origin.

Afterpulse measurements have been performed for the PMTs at room temperature and in LXe. Figure~\ref{Fig:Afterpulses_mxl} (right) shows a comparison of the afterpulse spectra for a PMT in each environment, according to the afterpulse groups defined in the left plot. An analysis of the afterpulse rates has been performed by recording several million single photon triggers (from dark pulses) and the pulses that follow within a 4\,$\mu$s window. The afterpulse rate is given as a percentage of the total SPE trigger signals. The mean afterpulse rate at room temperature for 44 PMTs is (1.4\,$\pm$\,1.2)\,\%. The ions producing the highest rates are He$^+$ (0.02\,$\pm$\,0.01)\,\%, CH$^+_4$ (0.07\,$\pm$\,0.03)\,\% and Ar$^+$ (0.08\,$\pm$\,0.07)\,\%. All other ions show rates below 0.02\,\% of the total trigger signals. The afterpulse rates were also recorded in LXe for 11 of those PMTs, in which case the mean afterpulse rate increased to (8.6\,$\pm$\,2.2)\,\%. The main increase was seen in the A2 group consisting of afterpulses with SPE amplitudes spread out homogeneously over several microseconds. These afterpulses may be caused by photons from the micro light emission described in section~\ref{sec:lightemission}. On the other hand, the average contributions of He$^+$ and Ar$^+$ remain unchanged, whereas the mean CH$^+_4$ rate decreased to (0.02\,$\pm$\,0.01)\,\%. This may be related to a stronger adhesion to the PMT surfaces for electronegative molecules (such as CH$_4$) than for noble elements (such as He and Ar). Upon warm-up to room temperature, the afterpulse behavior prior to cool-down is recovered, except in the case of the appearance of a Xe leak (which will be discussed in the following). The results of the measurements at room temperature and LXe are summarized in table~\ref{table:ap_rates}.

\begin{table}
\begin{centering}
\resizebox{\textwidth}{!}{%
\begin{tabular}{c|c|cc|ccc}
\hline
Afterpulses [\%]:  & Total & A1 & A2 & He$^+$ & CH$^+_4$ & Ar$^+$ \\ \hline
Room temp. & $1.4\pm1.2$ & $0.11\pm0.05$ & $0.9\pm1.1$ & $0.02\pm0.01$ & $0.07\pm\,0.03$ & $0.08\pm0.07$ \\ 
   LXe     & $8.6\pm2.2$ & $1.8\pm0.5$ & $6.4\pm1.7$ & $0.02\pm0.01$ & $0.02\pm0.01$ & $0.08\pm0.02$ \\
\hline 
\end{tabular}
}
\caption{Mean afterpulse rates given as a percentage of the total SPE trigger pulses. A comparison is made between the measurements at room temperature and in LXe. The uncertainties correspond to the mean absolute deviation of the distribution of afterpulse rates for the measured PMT sets. The rate increase in LXe is due to the A2 afterpulses from single photoelectrons distributed homogeneously in time, whereas the afterpulses from identified ions remain on average the same or decrease. Ions not shown in the table have an average rate below 0.02\,\% of the total triggers.}
 \label{table:ap_rates} 
\end{centering}
\end{table}

\begin{figure}[]
  \begin{center}
  \makebox[\textwidth][c]{
   \includegraphics[angle=0,width=0.52\textwidth]{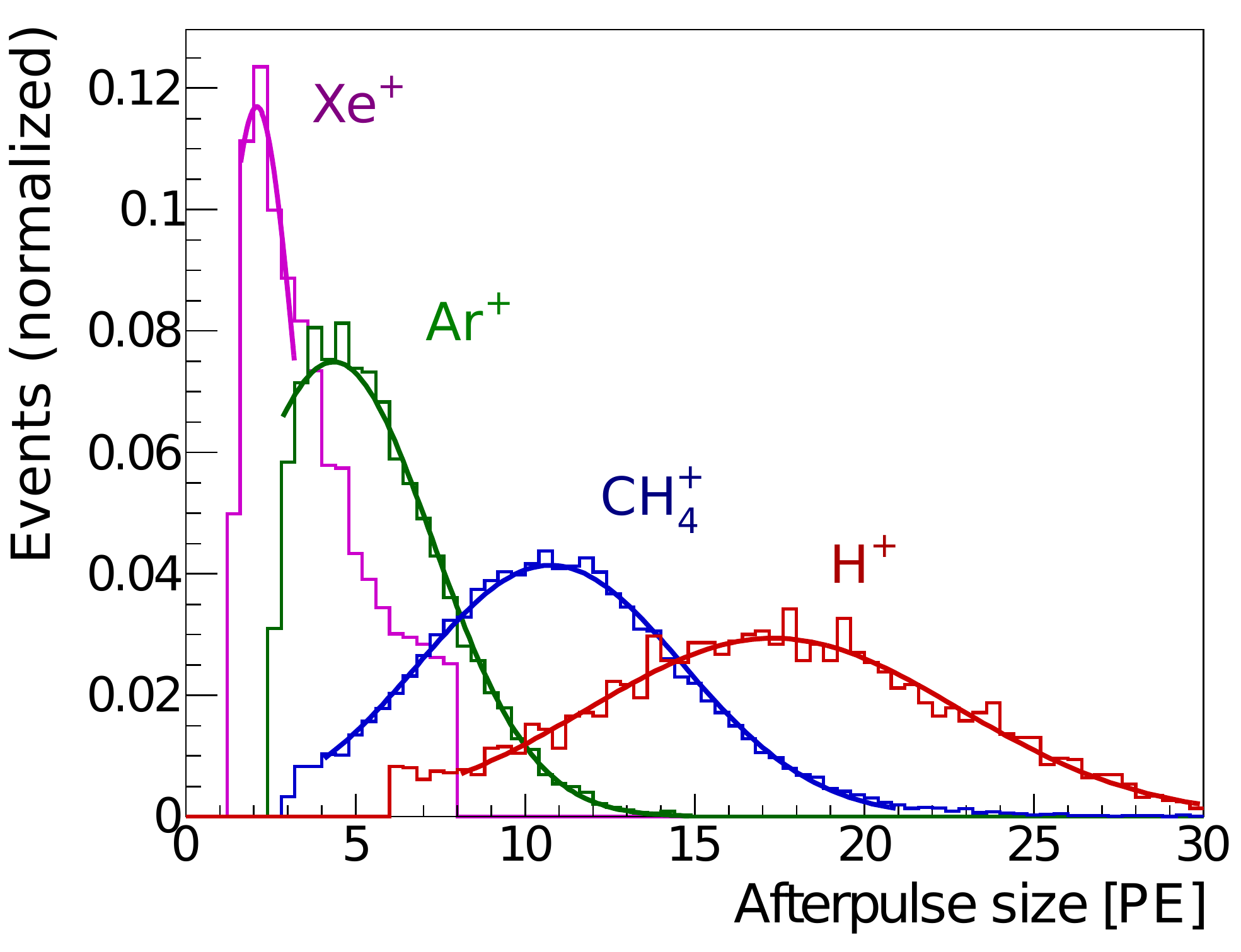}
   \includegraphics[angle=0,width=0.52\textwidth]{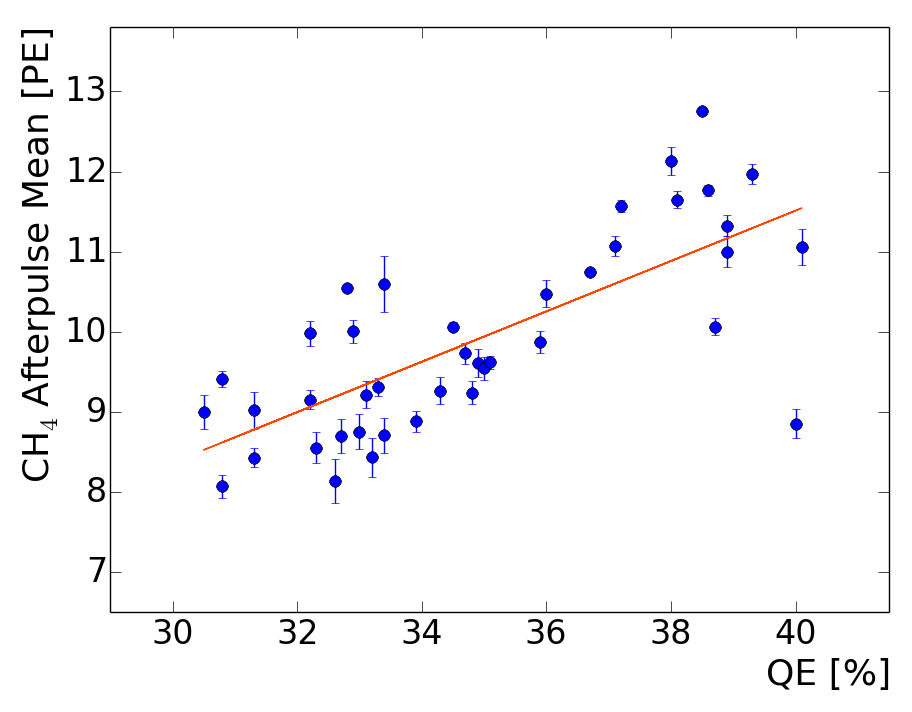}
   }
   \caption[]{(Left) Afterpulse size distributions for selected ions. Heavier ions like Xe$^+$ produce smaller afterpulses, while lighter ions like H$^+$, generate larger afterpulses. (Right) Distribution of the mean CH$^+_4$ afterpulse size compared to the QE of each PMT. A linear fit is applied to the data (orange line) and the Pearson correlation coefficient is found to be 0.74. This indicates a direct correlation between the observables, meaning that the measured number of photoelectrons in afterpulses gives an insight into the quantum efficiency of a PMT. }
   \label{Fig:ap_ions}
  \end{center}	
\end{figure}

The ions which most contribute to afterpulsing have been labeled in the example spectra of figure~\ref{Fig:Afterpulses_mxl}. It can be observed that the mean size of the afterpulses decreases with the mass of the ion. Figure~\ref{Fig:ap_ions} (left) shows the afterpulse size distribution for selected ions. Heavier ions like Xe$^+$ produce afterpulses with a mean around 2\,PE, while lighter ions like H$^+$ produce larger afterpulses of around 18\,PE with a larger spread. Figure~\ref{Fig:ap_ions} (right) shows the distribution of the mean CH$^+_4$ afterpulse size in PE compared to the QE of each PMT. The Pearson correlation coefficient is found to be 0.74, indicating a direct correlation between the observables. This shows that a photocathode with a higher electron yield from photons also generates more electrons from impinging ions. As a result, the measurements of afterpulse size in photoelectrons can give an insight into the quantum efficiency of a PMT.

\begin{figure}[]
  \begin{center}
   \includegraphics[angle=0,width=0.56\textwidth]{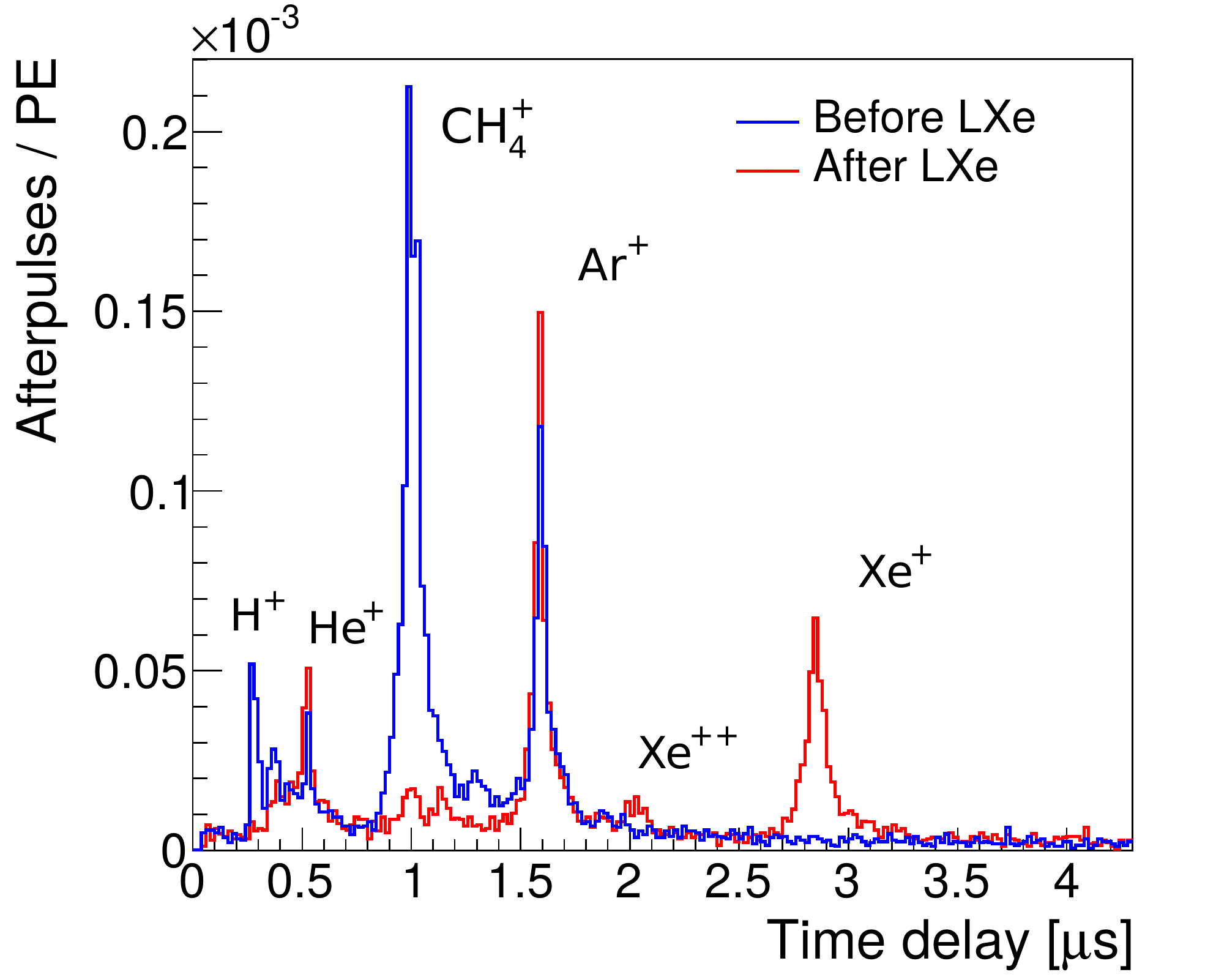}
   \caption[]{
   Example afterpulse spectra taken before (blue) and after (red) operation of a faulty PMT in liquid xenon. The appearance of a xenon peak at 2.8\,$\mu s$ is a clear indicator of a leak in the PMT. The decrease of CH$_4^+$ has been observed in all PMTs after operation at cryogenic temperatures, as presented in table\,\ref{table:ap_rates}.}
  \label{Fig:ap_leaks}
  \end{center}	
\end{figure}

This afterpulse analysis allows to diagnose the presence of leaks in faulty PMTs during cool down, as has also been done by other groups to identify Xe leaks (such as in~\cite{Li:2015qhq}). Out of the 44 PMTs tested in LXe, 8 presented the appearance of Xe afterpulses. Figure~\ref{Fig:ap_leaks} shows the spectra of the A3 group afterpulses for the same PMT before and after cool down. A clear rise of afterpulses around 2.8\,$\mu s$ is observed. For this PMT the Xe$^+$ afterpulse rate increased to 0.04\,\%, while the Xe$^+$ rates for other PMTs with a leak increased to as much as 0.98\,\%. These PMTs have been returned to the production company, where the presence of Xe in the tubes has been confirmed and the PMTs have been replaced.

It is important to remark that many of the PMTs tested in Xe were pre-selected from the observation of an N$_2$ afterpulse increase after their first cryogenic tests or other issues such as unstable DC rates or flashing (described in section~\ref{sec:lightemission}). Thus, the statistics for the appearance of Xe leaks might not be representative of the full set of PMTs installed in XENON1T. 

\section{Summary} \label{sec:sum}

In this article we report on the extensive testing campaign of 321 R11410-21 photomultiplier tubes, out of which 248 have been selected for long-term operation in the XENON1T experiment. The evaluation of the tubes is based on the dark count rate, SPE response, transit time spread, quantum efficiency, and level of light emission. Detailed measurements have been performed on a subset of PMTs in gaseous and liquid xenon. The evolution of the dark count rate and gain have been studied, showing the viability to operate the tubes long-term in a dark matter experiment. The identification of residual gas molecules within the PMTs through the analysis of afterpulses has been studied in detail, matching the experimental measurements with analytical calculations and simulations. The results have been used for diagnosis of the PMT vacuum quality and identification of leaks.

A total of 73 tubes, 22\,\% of the amount tested, have been rejected and thus excluded for use in XENON1T. Out of these, 12 tubes have been rejected due to an elevated or unstable dark count rate, either at cryogenic or room temperature. A larger amount, 53 in total, presented light emission. Out of the 44 PMTs tested in LXe, 8 were identified to show afterpulses from xenon ions, indicating a leak in the tube and thus being unsuited for long-term operation. It must be noted that the leak statistics is not representative of the whole set, since most PMTs tested in LXe were selected for their unstable performance during previous tests or suspicion of a leak after operation in nitrogen gas at $-100^{\circ}$C.   

The selected 248 PMTs have an average quantum efficiency of (34\,$\pm$\,3)\,\%  and show a low average dark count rate of (40\,$\pm$\,13)\,Hz at $-100\,^{\circ}$C.  
At a gain between $(2-3)\times 10^{6}$ the peak-to-valley ratio ranges from 2.5 to 4.5, proving the large signal to noise separation of this tube. During their operation in XENON1T, the evolution of the gain and dark count rates will be monitored for all PMTs. The methods presented here will also allow to identify the appearance of light emission or leaks in the tubes throughout the course of the experiment.

\section*{Acknowledgements}

We thank Hamamatsu for the fruitful 
collaboration and production of the PMTs used in these studies and in XENON1T. We also thank Andreas James, Daniel Florin and Martin Auger at UZH, and Reinhard Hofacker at MPIK, for their technical support. We gratefully acknowledge the support from the Swiss National Science Foundation, the European Union FP7 ITN INVISIBLES (Marie Curie Actions, PITN- GA-2011- 289442), the Max-Planck society and the DFG research training group 'Particle physics beyond the standard model'. We acknowledge the XENON Bologna colleagues for providing a few PMTs.

\bibliography{DMreview_v12}

\end{document}